\documentclass[12pt,emtex]{article}
\usepackage{amsfonts}
\usepackage{amssymb}
\usepackage{amscd}

\newcommand{\BEQ}{\begin{equation}}
\newcommand{\EEQ}{\end{equation}}

\newtheorem{Th}{Theorem}
\newtheorem{lem}{Lemma}
\newtheorem{Lem}{Lemma}
\newtheorem{Prop}{Proposition}

\newtheorem{Conj}{Conjecture}

\newtheorem{Def}{Definition}
\newtheorem{Rem}{Remark}

\def\nn{\nonumber}

\def\one#1{#1^{\raise5pt\hbox{$\scriptstyle\!\!\!\!1$}}\,{}}
\def\two#1{#1^{\raise5pt\hbox{$\scriptstyle\!\!\!\!2$}}\,{}}

\def\bea{\begin{eqnarray}}
\def\eea{\end{eqnarray}}

\def\C{{\mathbb{ C}}}
\def\A{{\mathbb{A}}}
\def\CC{{\mathbb{ C}}}
\def\RR{{\mathbb{ R}}}

\def\l{\lambda}

\def\g{\mathfrak{gl}_n}
\def\gg{\mathfrak{g}}
\def\p{\partial_z}
\def\h{{\mathsf h}}
\def\d{\displaystyle}

\def\pu{\partial_u}

\begin{document}

\begin{titlepage}
\hfill ITEP-TH-09/06 \vskip 1.2cm

\centerline{\LARGE Quantum spectral curves, }
\centerline{\LARGE quantum integrable systems}
\centerline{\LARGE and the geometric Langlands correspondence}

\vskip 1.1cm \centerline{A. Chervov \footnote{E-mail:
chervov@itep.ru} , D. Talalaev \footnote{E-mail:
talalaev@itep.ru} }

\centerline{\sf Institute for Theoretical and Experimental Physics
\footnote{ITEP, 25 B. Cheremushkinskaya, Moscow, 117259, Russia.}}

\vskip 1.0cm
\centerline{\large \bf Abstract}
\vskip 1cm
The spectral curve is the key ingredient in the modern theory
of classical integrable systems.
We develop a construction of the ``quantum spectral curve''
and argue that it takes the analogous structural and unifying role on the quantum level also.
In the simplest but essential case the ``quantum spectral curve''
is given by the formula $"det"(L(z)-\partial_z)$  [Talalaev04] (hep-th/0404153).

As an easy application of our constructions we obtain the following:
quite a universal recipe to define quantum commuting hamiltonians from classical ones,
in particular an
explicit description of a maximal commutative subalgebra in
$U( \g[t])/t^N$ and in
~$U(\g[t^{-1}])\otimes U(t\g[t])$;
the relation (isomorphism) of the
constructed commutative subalgebra with the center on the
critical level of $U( \hat \g)$ by the AKS-type arguments; an
explicit formula for the center generators and conjecture on $W$-algebra generators;
a recipe to obtain the $q$-deformation of these results;
a simple and explicit construction of the Langlands correspondence and
new points of view on its higher-dimensional generalization and relation
to "D-connections"; a relation between the ``quantum spectral curve''
and the Knizhnik-Zamolodchikov equation; new generalizations of the KZ-equation;
a conjecture on rationality of the solutions of the KZ-equation
for special values of level.

In simplest cases we observe coincidence of the
``quantum spectral curve'' and the so-called Baxter equation,
our results provide a general construction
of the Baxter equation. Connection with the KZ-equation
offers a new powerful way to construct the Baxter's $Q$-operator.
Generalizing the known observations on the connection
between the Baxter equation and the Bethe ansatz we formulate a conjecture
relating the spectrum of the underlying integrable model and
the properties of the ``quantum spectral curve''.
Our results are deeply related with the Sklyanin's approach
to separation of variables.

\vskip 1.0cm

\end{titlepage}

\newpage


\tableofcontents

\newpage

\section{Introduction and main results}

The spectral curve is the key ingredient in the modern theory
of classical integrable systems.
In this paper we argue that
the construction from \cite{Talalaev04} of ``quantum characteristic
polynomial'' (QCP) takes this structural and unifying role on the quantum level.
The generalizations
and applications were discussed in
\cite{CT04},
 \cite{MV2}.

One usually defines a spectral curve of a classical integrable model as $$det(L(z)-\l)=0$$
where $L(z)$ is the corresponding Lax operator. The moduli of such a curve
define conserved quantities, the dynamics is linearized on the Jacobian
(possibly generalized) of the spectral curve etc.
The ``quantum characteristic polynomial'' (QCP) in the simplest case can be
informally written as $$"det"(L(z)-\p)$$
There are another names for QCP -
"universal $G$-oper", "quantum spectral curve"
or "universal Baxter equation".
We show in this paper that
principal questions concerning quantum integrable systems
are intrinsically encoded in QCP.

We show the relevance of QCP
to the following questions:
\
\begin{enumerate}
  \item {\bf Quantization problem}
\\
QCP provides a universal recipe to construct quantum integrable
hamiltonian systems from classical ones. From the mathematical
point of view it gives a way to construct commutative subalgebras and central elements
in various associative algebras like universal enveloping algebras and quantum groups.

  \item {\bf Solution related ...}
\\
QCP provides a natural
generalization of the Baxter equation
and hence gives a way to find the spectrum of quantum integrable models.
With its help we relate the Baxter type equations to the
Knizhnik-Zamoldchikov type equations (see also \cite{CT04}),
giving a new
approach to the construction of the Baxter Q-operator which is
essential for finding the spectrum of quantum commuting hamiltonians
and other problems. The explanation of these constructions comes from Sklyanin's ideas
on separation of variables which are deeply related to the
"quantum characteristic polynomial".

\item {\bf Langlands correspondence}
\\
QCP with the AKS construction 
provides an elementary demonstration of the local
geometric Langlands correspondence over $\C$
 as a quite particular case of the general picture.
We shall show (see also \cite{CT06-1,CT06-2}) that it provides a simple
and explicit construction for the global geometric Langlands correspondence.

\end{enumerate}

We also remark on relations with the $D$-dimensional Langlands
correspondence ($D$-connections), Drinfeld-Sokolov reduction and
matrix models. In subsequent papers we apply the theorem above to quantize
the Hitchin system.

Let us briefly formulate the main constructions  of the paper
(see the main text for the notations and clarification).

\subsection{Main hero}

Let us present here the construction of the "quantum characteristic polynomial"
in the simplest but illustrative case. More general constructions are
described below.
\\
Consider an associative algebra $A_q$ and
$Mat_n\otimes A_q$-valued function  $L(z)$
(in application $L(z)$ will be
 the so-called Lax operator of some integrable system).
\\
{\Def~} The "quantum characteristic polynomial" QCP of $L(z)$
is an element of $A_q\otimes Diff[z]$, where $Diff[z]$ is the associative algebra
of differential operators in variable $z,$ defined by the following
expression:
\bea \label{qchar-intro-intro}
"det"(L(z)-\p)\stackrel{def}{=}Tr  A_n (L_1(z)-\p)\ldots
(L_n(z)-\p)=\sum_{k=0}^nC_n^k(-1)^{n-k}QI_k(z)\p^{n-k}
\eea
where $L_i(z)\in {Mat_n}^{\otimes n}\otimes
A_q((z))$ is obtained via the inclusion
$Mat_n\hookrightarrow
Mat_n^{\otimes n}$ as the $i$-th component. $A_n$ is an
antisymmetrization operator in ${\C^n}^{\otimes n},$ the trace is taken over
${Mat_n}^{\otimes n}.$

{\Rem ~} If one puts $\l$ instead of $\p$ in the formulas above and assumes
$A$ to be a commutative algebra then
one obtains the usual characteristic polynomial of $L(z)$ i.e. $det(L(z)-\l)$.

{\Rem ~} Actually the construction above gives a correct
definition only in the rational case. Proper modifications for
other cases are described below.

\subsection{Commutative subalgebra in $U( \g[t])/t^N$ }

{\bf Theorem } {\em
Consider the standard Lax operators for $U(\g[t])/t^N$
with $N$ finite or infinite:
\bea
\label{Lax-n1-intro}
L(z)=\sum_{i=0}^{N-1} \Phi_i z^{-(i+1)}
\eea
\\
Then the expressions $QI_k(z)$ given by Talalaev's quantum characteristic polynomial:
\bea \label{qchar-intro}
"det"(L(z)-\partial_z)\stackrel{def}{=}Tr A_n (L_1(z)-\p)\ldots
(L_n(z)-\p)=\sum_{k=0}^nC_n^k(-1)^{n-k}QI_k(z)\p^{n-k}
\eea
commute with each other
\bea
[QI_k(u),QI_m(v)]=0\nn
\eea
}
\\
{\Rem ~} Actually slightly more general theorem
 is proved below (see theorem
\ref{most-general-formulation}).

{\Rem~} One can see from the formula above (see \cite{Talalaev04})
or from the results of \cite{CRT04}
that the quadratic and cubic generators of this commutative subalgebra can
be described as $TrL^2(z), TrL^3(z)$ but this is not true for the higher
order generators: $TrL^4(z)$ does not commute with $Tr L^2(u).$

\subsection{The center of  $U_{crit} ( \widehat{\g})$ and W-algebras}

{\bf Theorem }{\em
Let $U_{crit} (\widehat{ \g})$ be the universal enveloping algebra of $\widehat{ \g}$
at the critical level,
denote by $I$ the isomorphism of vector spaces
$U(\g[t^{-1}])\otimes U(t~\g^{op}[t]) \to U_{crit} (\widehat{ \g})$ given by
$I(a\otimes b)=aS(b)$, where $S$ is the standard antipode.
Consider the standard Lax operators for $U(\g[t^{-1}])\otimes U(t\g^{op}[t])$
\bea
\label{Lax-n2-intro}
L^{full}(z)=\sum_{i=-\infty}^{+\infty} \Phi_i z^{-(i+1)}
\eea
\\
\begin{itemize}
\item
the expressions $QI_k(z)\in U(\g[t^{-1}])\otimes U(t\g^{op}[t])$
given by Talalaev's quantum characteristic polynomial:
\bea \label{qchar-intro2}
"det"(L^{full}(z)-\partial_z)\stackrel{def}{=}Tr A_n (L_1^{full}(z)-\p)\ldots
(L_n^{full}(z)-\p)=\nn\\
=\sum_{k=0}^nC_n^k(-1)^{n-k}QI_k(z)\p^{n-k}
\eea
commute with each other
\bea
[QI_k(u),QI_m(v)]=0\nn
\eea
\item The commutative subalgebra in $U(\g[t^{-1}])\otimes U(t\g^{op}[t])$
 generated by $QI_k(z)$ goes under the map  $I$ to the center of $U_{crit} (\widehat{ \g})$,
moreover this defines an isomorphism of commutative algebras.
\end{itemize}
Symbolically this can be written as follows:
\bea
:"det"(L_{full}(z)-\p) : ~~~\mbox{generates the center at the critical level}
\eea
}
\\
Here $:.. :$ stands for the normal ordering which should be understood as the map $I$.
This agrees with the standard prescription for the center.

{\bf Conjecture} {\em
\bea
:"det"(L_{full}(z)-\p) : ~~~\mbox{generates the W-algebra out of the critical level}
\eea
}

\subsection{Langlands correspondence}
Let us explain how our construction produces the local geometric Langlands correspondence over $\CC$
at the critical level
(the global case will be treated in subsequent publications \cite{CT06-1}, \cite{CT06-2}
in order to keep the volume of present).

The local geometric Langlands correspondence over $\CC$ at the critical level
aims to construct a bijection between representations of
$U_{crit}(\widehat{\g})$ (automorphic side) and connections on
$Spec~\CC((z))$ (Galois side). We approach this problem as follows:
representations are usually defined by fixing  values of Casimir
elements; it was proved in \cite{FeiginFrenkel} that the center
$Z(U_{crit}(\widehat{\g}))$ is huge enough (see also subsection above) and
for any character $\chi:Z(U_{crit}(\widehat{\g}))\to \CC$ there exists an
irreducible representation
$V_{\chi}$ of $U_{crit}(\widehat{\g})$ on which the center acts via $\chi$.
Now according to the theorem of the previous subsection one can
obtain the character
$$\chi_{M}: M_q\to \CC \qquad \mbox{ where} \qquad
M_q\subset U(\g[t^{-1}]\oplus t\g^{op}[t])$$
is the commutative subalgebra
defined with the help of $"det"(L^{full}(z)-\p)$. Now one obtains
a connection on  $Spec~\CC((z))$ in the following  way: one
considers the scalar differential operator of order $n$
\bea
\label{conn}
\chi_M["det"(L^{full}(z)-\p)]=\sum_{k=0}^nC_n^k(-1)^{n-k}\chi_M[QI_k(z)]\p^{n-k}
\eea
which defines a connection  on the trivial rank $n$ bundle ($GL(n)$-oper) over
$Spec~\CC((z))$.
\\
Schematically the desired correspondence looks like:
\bea
V_{\chi}\in Rep(U_{crit}(\widehat{\g})) \rightleftarrows
  \mbox{connection defined from~~}  \chi_M["det"(L^{full}(z)-\p)]\nn
\eea

\subsection{Separation of variables}

{\bf Conjecture.}
Consider a Poisson algebra $A,$ its center $Z$ and
its maximal Poisson commutative subalgebra $M$ given by
the spectral invariants of the Lax operator
(this means that there exists a $Mat_n\otimes A$-valued function\footnote{
rational for simplicity}  $L(z)$
such that:
$\forall z,\l~$  $det(L(z)-\l)$  belongs to $M$ and moreover $M$ is
generated by these values).
Denote also by $A_q$ the deformation quantization of $A$,
$Z_q$ - its center, $L_q(z)$ - the corresonding\footnote{Typically $L_q(z)=L(z)$}
Lax operator for $A_q$.
\\
{\bf Then}
$\exists z_i,\l_i$ in $A$ and $\hat z_i,\hat \l_i$ in $A_q$ called separated coordinates
\\
{\it
(more precisely these variables belong to a finite algebraic extension of the field of
fractions of $A$ and $A_q,$ this means that in general one cannot express these
variables in terms of rational functions of elements of $A,A_q$ but needs to
use algebraic functions i.e. to solve algebraic equations)}
\\
{\bf such  that}:
\begin{itemize}
\item $\{z_i,\l_j \}=\delta_{ij} f(z_i,\l_j) $ ~~~~~~~ $\{\l_i,\l_j \}=\{z_i,z_j \}=0$
\item $[\hat z_i,\hat \l_j ]=\delta_{ij} f_q(\hat z_i,\hat \l_j) $ ~~~~~~~
$[ \hat \l_i, \hat \l_j ]=[ \hat z_i,\hat z_j ]=0$
\item $z_i,\l_i$ and $Z$  are algebraically independent and  algebraically generate $A$\\
(analogously for $\hat z_i,\hat \l_i$,  $Z_q$ and  $A_q$)
\item  $\forall i~$   $det(L(z_i)-\l_i)=0 $ ~~~~~~ $\forall i~$   $"det"(L_q(\hat z_i)-\hat \l_i)=0 $
\item Let $\hat z, \hat \l$ be auxiliary variables (not elements of $A_q$) which
satisfy the  commutation relation $[\hat z,\hat \l ]=f_q(\hat z,\hat \l) $ (the same as for
quantum separated variables), then

$"det"(L_q(\hat z)-\hat \l)$ generates the commutative subalgebra in $A_q$ which quantizes subalgebra $M$.
\end{itemize}

Except for the last item these ideas are due to Sklyanin.
The last item assumes that the formula $"det"(L(z)-\p)$ is the simplest example
of the general construction given by $"det"(L(\hat z)-\hat \l)$.

{\Rem ~} This conjecture is mainly the belief that separation of
variables can be found for any reasonable integrable system.
If the parameter $z$ lives on some algebraic curve (not just $\C P^1$)
then $z,\l$ should be considered as generators of the ring of differential
operators on that curve
minus some  finite set of points.

\begin{itemize}
\item
Despite the fact that the change of variables to
separated ones is given by highly complicated algebraic functions
it is expected that
the generators of $M,M_q$ can be expressed via separated
variables by relatively simple
{\em rational} expressions analogous to those proposed in \cite{Babelon,EnriquezRubtsovSkewFields}.
\item
Classical separated variables can be found by the Sklyanin's "magic recipe":
\\
One should take the {\em properly} normalized Backer-Akhiezer function $\Phi(z)$,
i.e. the solution of $(L(z)-\l)\Phi(z)=0$
and consider the poles $z_i$ of this function together with  the values of $\l$
at these poles.
\end{itemize}

\subsection{Universal Baxter equation, $Q$-operator, $G$-opers, Bethe ansatz}
The Baxter equation (Baxter $T-Q$ relation) and the Baxter's $Q$-operator
are considered now as the most powerful  tools to find the spectrum of integrable models.
The known constructions are mostly restricted to the cases
of models related to $GL(2),~ GL(3).$
\\
Here we propose a general point of view on the Baxter equation and the $Q$-operator
and formulate a conjecture, which relates the Baxter equation to the spectrum
of an integrable model.
We explain that $G$-opers considered in the Langlands correspondence are
closely related to this subject.
\\
Let $A_q,M_q, \hat z,\hat \l$ be as in the previous subsection.
Consider $(\pi,V)$ - a representation of the algebra $A_q.$
Let $\hat z,\hat \l$  be realized as operators on
the space $\CC((z))$, where $\hat z $ acts as multiplication by $z$,
and $\hat \l$ acts as some differential (difference) operator $\tilde \l$.
\\
{\bf Construction of the Baxter equation.}
The differential (difference) equation
$$\pi("det"(L_q( z)-\tilde \l)) Q(z)=0$$
for an $End(V)$-valued function $Q(z)$ is called the Baxter equation and the
solution
$Q(z)$ - the Baxter's $Q$-operator.
\\
One can
see that for the $SU(2)$-Gaudin, $SU(2)$-XXX, XXZ models this equation
coincides with a degeneration
of the original Baxter equation.
\\
{\bf Main Conjecture}
Consider  a character $\chi: M_q\to \CC$ of the commutative algebra $M_q$,
consider the differential (difference) equation $\chi ("det"(L_q( z)-\tilde \l)) q(z)=0$,
then this equation has trivial monodromy ("$q$-monodromy" i.e. Birkhoff's connection
matrix \cite{Birk})
{\bf iff}
there exist: a unitary representation $(\pi,V)$ for $A_q$  and a joint
eigenvector $v$ for $M_q$
such that $\pi(m) v= \chi(m)v~~$  $\forall m\in M_q.$

~\\
{\em Historic remark.}
The conjecture above  was anticipated in the
literature. The origin of such claims goes back to R. Baxter, M. Gaudin, further
E. Sklyanin has related this to the separation of variables. Recently some
close results were obtained by E. Frenkel, E. Mukhin, V. Tarasov and A.
Varchenko \cite{FMTV}.
Our main source of knowledge and insparation on the subject
is \cite{Frenkel95} and the conjecture in this form is motivated by this paper.

~\\
{\em Corrections.} For the models related not to $GL(n)$ but to a
semisimple  group $G$ the  monodromy disappear only after passing to the  Langlands dual group $G^L$.
If irregular singularities appear in $\chi ("det"(L( z)-\tilde \l)) $
one should require the vanishing of the Stokes matrices.

~\\
{\em Simplification for compact groups.} In the case of the $U(n)$-Gaudin and $U(n)$-XXX models
the conjecture can be simplified requesting instead of absence
of monodromy  the rationality of all functions $q(z)$ satisfying the equation above.
In these cases  the conjecture should essentially  follow
(at least in one direction) from the results of
\cite{FMTV} (see also \cite{CT04}).

~\\
{\em Relation to G-opers.}
For the case $A_q=U(\g[t])$ the operator $\chi ("det"(L( z)-\tilde \l))$
is precisely the G-oper which was considered in the Langlands correspondence.

~\\
{\em Relation to Bethe equations.}
It is known that the condition of absence of monodromy can be
efficiently written down as the Bethe equations in the case of the Gaudin and XXX models.
In general an analogous procedure is not known.
However for example in the case of regular connections
the necessary condition for absence of monodromy
is the very trivial condition for the  residues  of
the connection to have integer spectrum.
Hopefully the Bethe equations for the spectrum is
the consequence of this similar local condition and
by some unknown reasons the absence of local
monodromies guarantees the  absence of the global one.
~\\
{\em "Bethe roots" are zeros of function $q(z)$.}
It is known that the Bethe equations are written on some
auxiliary variables $w_i$. It can be seen in examples that
$w_i$ are the zeroes of the function
$q(z)$ solving the equation  $\chi ("det"(L( z)-\tilde \l))q(z)=0$.
It is tempting to think that it is a general phenomenon.

~\\
{\em Explanations.} The explanation of the conjecture is obvious from the point of view
of separation of variables (see main text),
another explanation is related to an alternative  construction of the  Langlands
correspondence and relates it to the Bohr-Sommerfeld conditions (see \cite{CT06-2}).

\subsection{Knizhnik-Zamolodchikov and Baxter equations}

Let us preserve the notations from the subsection above.
\begin{itemize}
\item {\bf Observation} Let $A_q=U(\g\oplus...\oplus\g)$, $L^{G}(z)$ be
the standard Lax operator for the Gaudin model, let $(\pi,V)$ be a representation
of $A_q$, then
the equation on a $V\otimes \CC^n$-valued function $\Psi(z)$ of the type
$
(\p -\pi(L^G(z))) \Psi(z)=0
$
coincides precisely with one component of the standard system of rational KZ-equations.
\item {\bf Notations}
Let us take $A_q$, $L(z)$ as in the previous subsection and
$(\pi,V)$ - a representation of $A_q$,
let us call the equation on an $A_q\otimes \CC^n$-valued function $\Psi(z)$ of the type
$
( \tilde \l -L_q(z)) \Psi(z)=0
$
the {\em universal generalized KZ-equation}.

Let us call the equation on the $V\otimes \CC^n$-valued function $\Psi_\pi(z)$ of the type:\\
$
(\tilde \l -\pi(L_q(z))) \Psi_\pi(z)=0
$
the {\em generalized KZ-equation}.
\item {\bf KZ rationality conjecture}
The standard KZ-equation $
(\p -\pi(L^G(z))) \Psi(z)=0
$
has only rational solutions.
\item {\bf Generalized KZ rationality conjecture}
Consider $(\pi,V)$ - a unitary representation of $A_q$,
consider the equation $(\tilde \l - \pi (L_q(z))) \Psi_\pi(z)=0$
for a $V\otimes \CC^n$-valued function $\Psi_\pi(z)$.
Then this equation has trivial monodromy 
("$q$-monodromy" i.e. Birkhoff's connection matrix \cite{Birk}).
~\\
{\em Corrections.} For the models related not to $GL(n)$ but
to a semisimple  group $G$ the monodromy disappears only after passing to the Langlands dual group $G^L$.
If irregular singularities appear in $(\tilde \l - \pi (L_q(z)))$
one should require the vanishing of the Stokes matrices.

\item {\bf Baxter from KZ}
The main message of \cite{CT04} was the following:  one is able to construct
solutions of the Baxter equation from solutions of the KZ type equation.
This goes as follows:
let $\Psi(z)$ be a solution of the equation
$
(\tilde \l -L_q(z)) \Psi(z)=0
$.
Then any  component $\Psi_i(z)$ of the vector $\Psi(z)$
provides a solution of the universal Baxter equation:
$"det"(L( z)-\tilde \l)) \Psi_i(z)=0$.
\footnote{Moreover, in \cite{CT06-03} we show that  there is a gauge
transformation of the KZ-type connection $(\tilde \l -L_q(z))$
to the Drinfeld-Sokolov form corresponding to the differential (difference)
operator $"det"(L( z)-\tilde \l))$.}
\item
The construction above allows to deduce the main conjecture of the
 previous subsection at least
in one direction from the KZ-conjecture \cite{CT04}.
\end{itemize}

\vskip 1cm
\subsection{Acknowledgements}
The work of the both authors
has been partially supported by the RFBR grant 04-01-00702.
The work of one of the authors (AC) has been partially supported by the INTAS
grant YSF-04-83-3396,
the part of it was done during the visit to SISSA under the INTAS project,
the author is deeply indebted to SISSA and especially to G. Falqui
for providing warm hospitality, excellent working conditions and stimulating  discussions.
The authors are grateful to L.Rybnikov for explaining us the methods and results
of his paper prior to its publication.

\section{Commutative family in $U(\g[t])/t^N$ \label{Sect-main}}

In this section  we  show  how  to define a commutative subalgebra
in the algebra $U(\g[t])/(t^N=0)$ for $N=1,2,...,\infty$.
Moreover we show that it can be given exactly by the same explicit formula
as in \cite{Talalaev04}.
The motivation for the construction is quite straightforward:
the construction from \cite{Talalaev04} produces  the quantum commutative
subalgebra in $U( \g \oplus \g \oplus ... \oplus \g)$ which quantizes
the classical Gaudin hamiltonians. On the other hand
$$U( \g \oplus \g \oplus ... \oplus \g)= U(\g[t])/(P_N(t)=0)$$ where
$P_N(t)$ is a polynomial of degree $N$ with different roots.
Passing to the limit $P_N(t)\to t^N$ one can expect
to obtain a commutative subalgebra in $U(\g[t])/(t^N=0)$ of the same size.
Though passing to the limit is quite delicate operation
it is quite  natural that
there will be no problems in this situation because of the
general experience of the theory of integrable systems and results
of \cite{Chern,MussoIK},
where similar constructions have been realized in the classical case.
We give a direct, self-contained proof not appealing to other considerations.

\subsection{Main notation $"det"(L(z)-\partial_z)$}
In sections 2,3,4 we use the following notation.
Consider an associative algebra $A$, let $L(z)$ be an arbitrary  element in $
Mat_n\otimes A((z))$
(usually called the Lax operator),
then we denote by $"det"(L(z)-\partial_z)$  and call it the "quantum characteristic polynomial" of $L(z)$
the following expression
\bea
"det"(L(z)-\partial_z) \stackrel{def}{=}
Tr A_n (L_1(z)-\p)\ldots
(L_n(z)-\p)
\nn
\eea
We denote by $QI_k(z)$ the coefficients of this differential operator after
choosing an order: $\p^i$ on the right
\bea
\sum_{k=0}^nC_n^k(-1)^{n-k}QI_k(z)\p^{n-k}\stackrel{def}{=}Tr A_n (L_1(z)-\p)\ldots
(L_n(z)-\p)
\eea
In these formulas $L_i(z)$ is an
element of ${Mat_n}^{\otimes n}\otimes
A((z))$ obtained via the inclusion
$Mat_n\hookrightarrow
{Mat_n}^{\otimes n}$ as the $i$-th component. The element $A_n$ is an
antisymmetrization operator in ${\C^n}^{\otimes n}.$

\subsection{Quantization of the Gaudin model}
Let us recall the Talalaev's result \cite{Talalaev04}
on quantization of the Gaudin model \cite{Gaudin}.
The Lax operator is a rational function with  distinct poles:
\bea
\label{main-part-Lax-Gaud}
L^G(z)=K+\sum_{i=1...N}
\frac{\Phi^G_{i}}{z-z_i}\nn
\eea
where $K$ is an arbitrary constant matrix,
 $\Phi^G_i\in Mat_n\otimes \bigoplus\g \subset Mat_n\otimes
U(\g)^{\otimes N}$ is
 defined by the formula:
\bea \label{Phi-Gaudin}
\Phi^G_i=\sum_{kl}E_{kl}\otimes e_{kl}^{(i)}
\eea
where $E_{kl}$ form the  standard basis
in $Mat_n$ and $e_{kl}^{(i)}$ form a basis in the $i$-th copy of $\g.$
The quantum commutative family is constructed with the help of the quantum
characteristic polynomial
\bea \label{qchar}
"det"(L^G(z)-\partial_z)=Tr A_n (L^G_1(z)-\p)\ldots
(L^G_n(z)-\p)=\sum_{k=0}^nC_n^k(-1)^{n-k}QI_k(z)\p^{n-k}
\eea
{\bf Theorem \cite{Talalaev04}} the coefficients $QI_k(z)$ commute
\bea
[QI_k(z),QI_m(u)]=0\nn
\eea
and quantize the classical Gaudin hamiltonians.


{\Rem ~} Actually in \cite{Talalaev04} the case $K=0$ was considered. The case $K\ne0$
is similar and can be found in the appendix to the present paper.

\subsection{Commutative family in $U(\g[t])/t^N$ }
\paragraph{Family of structures}~
\\
Let us recall some standard facts about $r$-matrix and bihamiltonian
structures (see \cite{Hurtubrise,FalquiMusso,ReymanSemenov}).
Actually these papers discuss only the Poisson structures but
there is really no difference in our situation between classical and
quantum cases.
\\
For a polynomial $P_N(t)$ of degree $N$
the vector space $U(\g[t])/(P_N(t)=0)$ can be identified with the
space $U(\g[t])/(t^N=0)$. Hence consideration of different $P_N(t)$
gives different algebraic structures on the same vector space.
Let us recall how to write these algebraic structures with the help
of $r$-matrices. Consider the standard Lax operator for $U(\g[t])/(t^N=0)$
and the standard Lax operator for $U(\g[t])/(t^N=0)$
with a constant term $K$:
\bea
\label{Lax-n}
L(z)=\sum_{i=0}^{N-1} \Phi_i z^{-(i+1)} ~~~~~~~
L_K(z)=K+\sum_{i=0}^{N-1} \Phi_i z^{-(i+1)}
\eea
Here $K$ is an arbitrary constant matrix,
$z$ - a formal parameter (not the same as $t$ !),
$$\Phi_i\in \Bigl( Mat_n\otimes U(\g[t])/(t^N=0) \Bigr) \cong
\Bigl( Mat_n[U(\g[t])/(t^N=0)] \Bigr)$$
are given by
\bea
\label{not1}
\Phi_i=\sum_{kl} E_{kl}\otimes e_{kl}t^i \Longleftrightarrow (\Phi_i)_{kl}=e_{kl}t^i
\eea
where
$ E_{kl}\in Mat_n$,  $e_{kl}t^i\in U(\g[t]).$ Both $E_{kl}$ and $e_{kl}$
are the matrices with the only $(k,l)$-th nontrivial matrix element equal
to $1.$ We consider them as elements of different
algebras: the first is an element of the associative algebra $Mat_n$, the second -
of the universal enveloping algebra $U(\g[t])$.

Let us consider the standard $r$-matrix notations for the commutators in presence
of an arbitrary polynomial $f(z)$ of degree less or equal then $N:$
\bea
\label{rel1}
[\one L(z) , \two L(u) ] = [ \frac{P}{z-u}, \one L(z) \frac{f(u)}{u^N}
+ \two L(u) \frac{f(z)}{z^N} ]
\eea

Recall the following well-known simple fact:

{\Prop ~ \label{r-matr-comrelProp}
\begin{itemize}
  \item For $f(z)=z^N$  the algebraic structure defined by the relation
(\ref{rel1}) on the vector
space $U(\g[t])/(t^N=0)$ coincides with the standard algebraic structure on
$U(\g[t])/(t^n=0)$ and the isomorphism is identical:  $\Phi_i\to\Phi_i$.
  \item  For the generic  polynomial with distinct roots  $f(z)=\prod_{i=1}^N (z-z_i)$
the algebra defined on the vector space $U(\g[t])/(t^N=0)$
is isomorphic to $U( \g \oplus \g \oplus ... \oplus \g)$
and the isomorphism $(\Phi_0,...,\Phi_{N-1}) \to  (\Phi^G_0,..., \Phi^G_{N-1})$
is given by the decomposition to simple fractions of:
$$L(z)z^N/f(z)=\sum_i \frac {\Phi^G_i}{z-z_i}$$
\end{itemize}
}
{\Rem ~} For an arbitrary  polynomial  $f(z)$ with possibly multiple roots
the algebra defined on the vector space $U(\g[t])/(t^N=0)$
is isomorphic to $U( \g[t])/(f(t)=0)$
and the isomorphism
is defined by the decomposition to simple fractions with possibly higher order poles of
$L(z)z^N/f(z)$ in an obvious way.
{\Rem ~} In the classical case the formula above defines a pencil
of compatible Poisson structures parameterized by $f(z)$.

\newpage
\paragraph{Commutative family}~
{\Th ~ \label{MainTh}
The Talalaev's formula
\bea
"det"(L_K(z)-\partial_z)
\eea
defines a commutative subalgebra in $U(\g[t])/(t^N=0)$.
\label{gen}
The subalgebra is maximal for $K$ with distinct eigenvalues.
}
\\
{\bf Proof~}
Let us start with the case $K=0.$
According to the result of \cite{Talalaev04} the coefficients of
$"det"(L^G(z)-\partial_z)$ generate a commutative subalgebra in
$U( \g \oplus \g \oplus ... \oplus \g)$ where $ L^G(z) $ is the Gaudin
Lax operator.
In our situation $L(z) z^N/f(z)$ is of the Gaudin type if $f(z)$ has
distinct roots (see Proposition 1)
and hence
the expression $"det"( L(z) z^N/f(z) -\partial_z)$
gives a commutative subalgebra in $U(\g[t])/(f(t)=0)$ for such $f(z)$.
We have to analyze the properties of the  limit $f(t)\to t^N$ to prove the result.
\\
The fact that the constructed family stays commutative and does not
degenerate  in the limit is evident in a sense due to the following
arguments:
\begin{enumerate}
  \item The considered family of algebras is a family of universal enveloping
algebras for the family of Lie algebras constructed on the same vector space
by relations with different structure constants. This means that one has the
same Poincar\'e-Birkhof-Witt basis for the whole family of algebras.
  \item The structure constants depend on coefficients of the function $f(z)$
polynomially.
  \item The quantities $QI_i(z)$  being expressed in
the common PBW basis have coefficients which depend rationally on
coefficients of $f(z).$
  \item At generic points and at the point $f(z)=z^N$ the expressions $QI_i(z)$ are well
defined (they are not defined only at $f(z)=0$).
  \item The commutators $[QI_i(z),QI_j(u)]$ are rational functions on
coefficients of $f(z)$ on a such open set where $QI_i(z)$ are well
defined.
\end{enumerate}
The proposition now follows from the fact that the quantities $QI_i(z)$ are
well defined at the point $f(z)=z^N$ and the fact that there is
only the zero rational function which vanishes on an open set.
\\
Let us remark that the size of the commutative algebra remains the same
after passing to the limit $f(z)\to z^N.$ Consider the classical limit which
is the projection to the associated graded algebra in this case. The
generators $QI_i(z)$ map to the coefficients of the classical characteristic
polynomial $I_i(z).$ For all $f(z)$ the Poisson commutative subalgebra generated by the
coefficients of the expansion of $I_i(z)$ at poles  remains of the same
dimension  which is well known.  This implies the same fact on the quantum
level.
\\
The case $K\ne 0$ goes through the same lines because the quantization of the Gaudin
model with the free term is also valid (see appendices) $\blacksquare$

{\Rem ~}
One also has to note that the obtained subalgebra
 is maximal for $K$ with distinct eigenvalues due to maximality of the
Bethe subalgebra \cite{NO}.

\subsection{Commutative family in $U(\g[t])$ \label{sect-Ninfty}}
{\Prop ~ The commutative subalgebra in $U(\g[t])$ is given by the same formula
as in the case of truncated algebra
\bea
"det"(L(z)-\partial_z)
\eea
where $L(z)=\sum_{i=0}^{\infty} \Phi_i z^{-(i+1)}$.
}~
\\
This is a straightforward corollary of Theorem \ref{gen}.
The proof follows from the  consideration of the natural projections
$$U(\g[t])\to U(\g[t]/t^n)$$ Supposing that an element in $U(\g[t])$ is nonzero one
can always find such $n$ that the projection to $U(\g[t]/t^n)$ is also
nonzero. From this one sees that commutativity in $U(\g[t])$
is a corollary of commutativity in  $U(\g[t]/t^n)$.

\subsection{Historic remarks}
{\Rem~} The classical analogue  of the formula
$"det"(L(z)-\p)$ is just the characteristic polynomial
$det(L(z)-\l)
$.
It follows easily from the $r$-matrix technique that it provides classical
commuting expressions in  $S(\g[t])/t^N.$ The commuting quantities obtained in
the
way described above give rise to the analogue of the Gaudin integrable system
well-known for a long time \cite{ReymanSemenovFrenkel80,AdlerMoerbek80}.

{\Rem~} The central elements in $S(\g[t])/(t^N)$ ~~(~$U(\g[t])/(t^N)$~)
have been described in \cite{RaisTauvel92} (respectively \cite{Molev}).
Our subalgebra can be easily extended to a maximal subalgebra (see main
text).
Molev gave an explicit formula for the generators of center as
$\widetilde{det} \widetilde{L(z)}$ where his $\widetilde{det}$
and  $\widetilde{L(z)}$ are slightly different from the our's, it would
be interesting to clarify the relation.

{\Rem~} The considered commutative subalgebra in $U(\g[t])$
is in a sense a degeneration of the commutative subalgebra
in Yangian $Y(\g[t])$ (called the Bethe subalgebra in \cite{NO}) which seems to be discovered by the Faddeev's school
(see ex. \cite{Skl95SV} formula 3.16) but we are unable to find the precise reference.
In the mathematical literature it has been analyzed first in
 \cite{NO},
where all the proofs are provided and the generalization to
semisimple Lie algebras found.
The work \cite{Talalaev04} is heavily based on this paper.

\subsection{The most general formulation}
{\Th \label{most-general-formulation}
Consider an associative algebra $A$ and
a slightly more general Lax operator $L(z)\in A\otimes Mat_n((z))$:
\bea
\label{Lax-A-proofSect}
L(z)=\sum_{i=-M}^L \Phi_i z^{-(i+1)}
\eea
where $M\ge 0,L>0$ (one of $M,L$ can be infinite,
for the case $M=L=\infty$ see proposition below)
\\
such that this operator satisfy the commutation relation:
\bea
\label{rel-GaudA-proofSect}
[\one L(z) , \two L(u) ] = [ \frac{P}{z-u}, \one L(z)  + \two L(u)  ]
\eea
Then the expressions $QI_k(z)$ given by:
\bea \label{qchar-proofSect-A}
"det"(L(z)-\partial_z)=Tr A_n ({L}_1(z)-\p)\ldots
({L}_n(z)-\p)=\sum_{k=0}^nC_n^k(-1)^{n-k}QI_k(z)\p^{n-k}
\eea
commute with each other
\bea
[QI_k(u),QI_m(v)]=0\nn
\eea
}
\\~
{\bf Proof~} This follows from the fact that
the commutation relation \ref{rel-GaudA-proofSect}
implies that the algebra generated by $\Phi_{i,kl}$ is a factor algebra
of $U(\g[t]/t^M \oplus t~\g^{op}[t]/t^L)$.
The proof of the result for the algebra $U(\g[t]/t^M \oplus t~\g^{op}[t]/t^L)$
follows from the  proof of the  Theorem \ref{gen}
for the case $N=M+L$. The only change in the proof is that one needs to take the limit $f(z)\to z^{M+1}$
(not $f(z)\to z^{N}$).
\\
The case $M=\infty$ xor $N=\infty$ can be obtained
by the same arguments as in section \ref{sect-Ninfty}.
\\
For the case $M=\infty$ and  $N=\infty$ the theorem is not directly
applicable because it seems to be not true that
commutation relation \ref{rel-GaudA-proofSect}
implies that $A$ is a factor algebra
of $U(\g[t] \oplus t~\g^{op}[t])$.
All other arguments works without any change
$\square$

~
\\
So we obtain the following:
{\Prop If $M=\infty$, $N=\infty$ and the
algebra generated by $\Phi_{i,kl}$ is a factor algebra
of $U(\g[t] \oplus t~\g^{op}[t])$,
then the proposition above is also true.}
\\
This proposition for the case $A=U(\g[t] \oplus t~\g^{op}[t])$
will be essential in the next sections.

\subsection{Remark on the $GL$-invariance \label{GL-invar-sect}}
It is quite clear that:

{\Prop The commutative subalgebra obtained by $"det"(L(z)-\p)$
is invariant with respect to the action
of $\g\in \g[t,t^{-1}]$.
}

{\Rem ~} This is analogous to the fact that the Gaudin's hamiltonians are invariant
with respect to the diagonal action of $GL_n.$
\\
Let us remark that instead of introducing $K$ into the Lax operator
to produce a maximal commutative subalgebra
one can add to our subalgebra the generators of a maximal commutative subalgebra
constructed
from $\g$, for example of the Gelfand-Zeitlin subalgebra
(see \cite{FalquiMusso2} for the Gaudin case).

\subsection{Quantization of the "argument translation" method}

For the case $\g=\g[t]/t$ the Lax operator is simply $L_{MF}(z)=K+\Phi/z$.
The classical commutative subalgebra in $S(\g)$ obtained from coefficients of
$det(L_{MF}(z)-\l)$
was called Mishenko-Fomenko subalgebra in \cite{Vinberg}.
It originates from Manakov's paper \cite{Manakov} and was
deeply developed by Mishenko, Fomenko and their school (see \cite{Mish-F}
and subsequent papers)
under the name of the "argument translation method" (see also \cite{HarnadOnMF}).

~\\
From the results above it is cleat that:
{\Prop The Talalaev's formula
$
"det"(L_{MF}(z)-\partial_z)
$
defines a commutative subalgebra in $U(\g)$ that quantizes the MF-subalgebra.
}
{\Rem ~} {\em Relation with the other constructions}.
The formula above gives an explicit recipe for quantization. By \cite{Tarasov2}
the quantization is unique in this situation and hence coincides with the previous results
\cite{Tarasov1},\cite{NO}. It was conjectured in \cite{Vinberg} and proved in \cite{Tarasov1}
that the symmetrization
map $S(\g)\to U(\g)$ produces a commutative subalgebra in $U(\g)$ from
the MF-subalgebra in $S(\g)$).
The low degree hamiltonians can be constructed as $Tr L_{MF}(z)^k$ for $k<6$,
but $Tr L_{MF}(z)^6$ does not commute with $Tr L_{MF}(z)^3$
(\cite{CRT04}).

{\Rem ~} {\em Limit to Gelfand-Zeitlin}. It was shown in \cite{Vinberg} that
the Gelfand-Zeitlin subalgebra
in $U(\g)$ can be obtained as a limit of the MF-subalgebras.
It would be very interesting to understand the behaviour
of our construction in this limit
and the relation with deep results of \cite{GKL}
on the quantum Gelfand-Zeitlin integrable system.

\subsection{Coordinate change in QCP}

Consider the Lie algebra $t \g[t]/t^{N+1}$ and
the  Lax operator:
\bea
L(z)=\sum_{i=1}^{N} \Phi_i/ z^{i+1}
\eea
where $\Phi_i$ are given by the formula
\ref{not1}.
\\~\\
In our study of Hitchin system the following proposition will be used:

{\Th ~ The Talalaev's formula
\bea
"det"(L(z)-\partial_z)
\eea
defines a commutative subalgebra in $U(t\g[t])/(t^{N+1})$
}~

{\Rem ~} The proof is given in Appendix B.
Let us only make a comment that this theorem does not follow directly from theorem
\ref{most-general-formulation} since this Lax operator does not
satisfy the basic commutation relation:
\bea
\label{rel1-L+}
[\one L(z) , \two L(u) ] \ne  [ \frac{P}{z-u}, \one L(z) + \two L(u) ]
\eea
\\
Actually the following more general fact was proved:

{\Prop ~
The following invariant formula
$$"det"( L(z) dz-d^{deRham})$$
defines the
same commutative subalgebra for any change of variables $z=f(z')$.}
\\
We see that the quantum Lax operator
can be naturally interpreted as a connection, this is prompting
for higher-dimensional generalizations (see below).


\section{The center of $U(\widehat{ \g})$ at the critical level \label{critCent-sect}}

Let us explain the coincidence of the commutative subalgebras
obtained above with the ones
constructed from the center
of $U_{crit}(\widehat{\g})$ (\cite{FeiginFrenkel}).
Some related considerations for the Gaudin model can be found in
\cite{ER1,Semenov97}.

\subsection{Commutative subalgebras from the  center}

The idea to obtain a commutative subalgebra from the center
is due to Adler, Kostant and Symes \cite{AKS7980} and \cite{LebedevManin79}.
It was deeply developed by
Reyman and Semenov-Tian-Shansky. Let us recall the relevant material
on AKS constructions following \cite{Semenov97}.

Let $\gg=\gg_+\oplus \gg_-$ be a finite dimensional Lie algebra which
is a direct sum of two its Lie subalgebras.
Let one introduce a new Lie algebra structure on $\gg$:
 $[(h_1,f_1),  (h_2,f_2)]= ([ h_1, h_2],  -[f_1,f_2])$ where
$(h_i,f_i)\in \gg,~h_i\in \gg_+,~f_i\in \gg_-$.
Denote $\gg$ with the new bracket by $\gg_r$.

One has two isomorphisms of linear spaces 
\bea
\phi_{cl}&:&S(\gg) \rightarrow S(\gg_+)\otimes S(\gg_-^{op})\simeq
S(\gg_r)\nn\\
\phi &:& U(\gg) \rightarrow U(\gg_+)\otimes U(\gg_-^{op})\simeq U(\gg_r)\nn
\eea
the inverse maps are given by:
\bea
\phi_{cl}^{-1} (a\otimes b) = as(b), ~~~~~ \phi^{-1} (a\otimes b) = as(b)
\eea
where $s$ is the canonical antipode.

{\lem The center of the Poisson algebra $Z_{cl}\subset S(\gg)$
maps to a commutative subalgebra
in $S(\gg_+)\otimes S(\gg_-^{op})$ via $\phi_{cl}$ where
$\gg_-^{op}$ is $\gg_-$ with inverted bracket.
}
{\lem The center $Z\subset U(\gg)$ maps to a commutative subalgebra
in $U(\gg_+)\otimes U(\gg_-^{op})$ via $\phi.$
}
\\
For convenience we propose a proof of these classical facts in appendix C.
{\Rem~ \label{grading}} To use this result in infinite dimensional case one has to consider an
appropriate completion, for the case of $U_{crit}(\widehat{\g})$ it is sufficient to
consider a bigrading $deg(gt^k)=(k,0),~deg(gt^{-k})=(0,k)$ for $k\geq 0.$
The only thing to prove is that the central elements are elements of the
considered completion $U_{crit}^\cdot(\widehat{\g}).$ This is indeed the case due to
the arguments of the classical limit. In what follows we omit the dot for
the completed algebras $U_{crit}(\widehat{\g}),~U(\gg_r),$ corresponding
tensor products etc.

~\\
There is another canonical isomorphism of linear spaces
$$\varsigma:U(\gg)\rightarrow U(\gg_+)\oplus U(\gg)\gg_-$$
Let $\varphi$ be the projection onto the first summand $U(\gg_+).$
{\lem The center $Z\subset U(\gg)$ maps to a commutative subalgebra in
$U(\gg_+)$ via $\varphi.$}
\\
{\bf Proof~} Let $c_1,c_2\in Z.$
\bea
[c_1-\varphi(c_1),c_2-\varphi(c_2)]=[\varphi(c_1),\varphi(c_2)]\nn
\eea
r.h.s. $\in U(\gg_+);$ l.h.s. $\in U(\gg)\gg_-;$ hence both are zeroes
$\square$

~
\\
For applying this to current algebras let us firstly recall the following

{\Prop \label{gr-current}
Consider $\gg=\g[t,t^{-1}]=\g[t^{-1}]\oplus t\g[t]$
$$L_{full}(z)=\sum_{i=-\infty,\infty} \Phi_i z^{-i-1}$$
then the Lie algebra structure $g_r$ can be described by the commutation
relation
\bea
\label{rel-Gaud-center-sect}
[\one L_{full}(z) , \two L_{full}(u) ] = [ \frac{P}{z-u}, \one L_{full}(z)  + \two L_{full}(u)  ]
\eea
}

\subsection{$Z(U_{crit}(\widehat{\g}))$
and the commutative subalgebra in $U(t\g[t])$ }

Let us denote by $U_{crit}(\widehat{\g})$ the algebra
$U(\widehat{\g})$ at the critical level (for details see
\cite{FeiginFrenkel} where it was proved that $U_{crit}(\widehat{\g})$  has big center).

It also follows from this paper that the AKS construction works in this situation,
i.e. the natural projection map
$\varphi: U_{crit}(\widehat{\g})) \to U(t\g[t])$, restricted to $Z(U_{crit}(\widehat{\g}))$
produces a huge commutative subalgebra in $U(t\g[t])$.

{\Th The commutative subalgebra in  $U(t\g[t])$ defined in the present paper
with the help of  $"det"(L(z)-\p)$ coincides
with the subalgebra obtained from the center of $U_{crit}(\widehat{ \g})$
by the natural projection $\varphi: U_{crit}(\widehat{ \g}) \to U(\g[t])$.
}
\\
{\bf Proof~} The proof is based on \cite{RybnikovNew} where it was proved
that the centralizer in $U( t\g[t])$ of the quadratic Gaudin's hamiltonian $H_2$ is the
commutative subalgebra obtained from the center of $U(\widehat{ \g})$ at the critical level.
It is quite surprising that such a huge subalgebra is defined only by
one element,
similar results are known for Mishenko-Fomenko subalgebra \cite{Tarasov2}, \cite{Rybnikov05},
 Calogero hamiltonians \cite{OOS}.
\\
Since the commutative subalgebra given by $"det"(L(z)-\p)$ commutes
with $H_2$ it is a subset of the algebra obtained from the center. The
maximality property provides that
they coincide.
Instead of appealing to maximality one can say that the classical versions
of both subalgebras coincide and hence they have the same size
$\square$
{\Rem~}
The same proof works in the case of the projection to $U( \g[t]).$ One has
just to take into account that both subalgebras are invariant with respect to the global
$GL(n)$ action (see subsection \ref{GL-invar-sect}).

{\Rem ~} The AKS construction can be modified to incorporate the Lax
operator with the free term
$K$ (see \cite{HarnadOnMF}). The subalgebra
defined with the help of this modified construction should coincide with
the subalgebra obtained from $"det"(L_K(z)-\p)$.

\subsection{Explicit description of $Z(U_{crit}(\widehat{\g}))$
and W-algebras. }
Let us introduce the following notation $L_{full}(z)=\sum_{i=-\infty,\infty} \Phi_i z^{-i-1}.$
{\Th \label{center}
The center of $U_{crit}(\widehat{\g})$ is isomorphic as a commutative algebra
to the commutative subalgebra in $U(\g[t^{-1}]\oplus t\g^{op}[t])$ defined by
the Talalaev's formula $"det"(L_{full}(z)-\p)$. The isomorphism
is given by the map
\bea
I : U(\g[t^{-1}])\otimes U(t\g^{op}[t]) \to U_{crit}(\widehat{\g}),
\qquad
I : h_1 \otimes h_2 \to h_1s(h_2)
\eea
where $s$ is the standard antipode: $s(h)=-h$ on generators.}
\\
Symbolically this can be written as follows:
\bea
:"det"(L_{full}(z)-\p) : ~~~\mbox{generates the center at the critical level}
\eea
\\
Here $:.. :$ stands for the normal ordering which should be understood as the map $I$.
This construction agrees with the standard prescription for the center.

{\Rem ~} For the quadratic elements this was known  (Sugavara's formula).

{\Conj
\bea
:"det"(L_{full}(z)-\p) : ~~~\mbox{generates the W algebra out of the critical level}
\eea
}
\\
{\bf Proof of theorem \ref{center}~}
The proof is based on the same strategy as in \cite{RybnikovNew}, namely we
first prove that the algebra generated by the coefficients of the ``quantum characteristic
polynomial'' (QCP) of $L_{full}(z)$ coincides with the centralizer of the quadratic
coefficients and further use the Sugavara formula for the quadratic central
elements to prove that its images in $U(\g[t^{-1})\otimes U(t\g[t])$ coincide with the
quadratic coefficients of QCP. To prove the first statement we use a special
limit of the commutative family.
\\
For the reader's convenience let us recall once more the structure relations
\bea
[\one L_{full}(z), \two L_{full}(u)] = [\one L_{full}(z),P \delta(z-u)] \mbox{~~in~ } \g[t,t^{-1}]
\\ ~
 ~ [ \one L_{full}(z), \two L_{full}(u)] = [\frac{P }{z-u}, \one L_{full}(z)+ \two L_{full}(u)]
\mbox{~~in ~ }
\g[t^{-1}]\oplus t\g^{op}[t]
\eea
The Lie algebra $\g[t^{-1}]\oplus t\g^{op}[t]$ here is the $r$-matrix
Lie algebra  $g_r$ for $g=\g[t,t^{-1}]$ defined by the
decomposition $\g[t,t^{-1}]=\g[t^{-1}]\oplus t\g^{op}[t]$.
From the commutation relations above and standard $r$-matrix computations
one obtains that $Tr L_{full}^m(z)$ are central in $S\g[t,t^{-1}]$
and  $Tr L_{full}^m(z)$ generate a Poisson commutative subalegbra in
$S(\g[t^{-1}]\oplus t\g^{op}[t])$.

~\\
Let us consider a family of automorphisms of the algebra $U(\g[t^{-1}])\otimes U(t\g^{op}[t])$
defined in terms of the Lax operator as follows: let $K$ be a generic
diagonal  $n\times n$ matrix, the Lax operator
$$L_{full}^{\hbar}(z)=L_{full}(z)+\hbar K$$ satisfy the same $r$-matrix relation
as $L_{full}(z),$ hence this transformation provides a family of automorphisms depending on a
parameter $\hbar.$
\\
One has also a family of commutative subalgebras $M^\hbar$ in $U(\g[t^{-1}])\otimes U(t\g^{op}[t])$
defined by $"det"(L_{full}^{\hbar}(z)-\p).$ $M^\hbar$ centralizes the quadratic hamiltonian
$QI_2(L_{full}^{\hbar}(z)).$ The $k$-th generator $QI_k(z,\hbar)$ has the following scalar
leading term in $\hbar$
$$QI_k(z,\hbar)=\hbar^k TrA_n K_1K_2\ldots K_k + O(\hbar^{k-1})$$ We prefer
to
slightly change the basis of generators
$$QI_k(z,\hbar)\mapsto
\widetilde{QI_k}(z,\hbar)=(QI_k(z,\hbar)-\hbar^k TrA_n K_1K_2\ldots K_k)\hbar^{-k+1}$$
and consider the limit $\hbar\rightarrow\infty$
\bea
\widetilde{QI_k}(z,\hbar)\rightarrow Tr(L_{full}(z)K^{k-1})\nn
\eea
In this limit the considered expressions
generate the Cartan subalgebra $\mathfrak{H}=
\mathfrak{H}_-\otimes \mathfrak{H}_+=U(\mathfrak{h}[t^{-1}])\otimes
U(t\mathfrak{h}[t])$
Let us demonstrate that this subalgebra coincides with the centralizer of
$$ H_2^{\infty}(z)=lim_{\hbar\rightarrow \infty}\widetilde{QI}_2(z,\hbar)
=\sum_{i=-\infty,\infty} Tr(\Phi_iK) z^{-i-1}$$
{\Rem ~} All over this subsection the completion subject to the bigrading of
Remark \ref{grading} is meant.

~\\
It is obvious that $\mathfrak{H}\subset C(H_2^{\infty}(z)).$ Let the diagonal elements
of $K$ be $(k_1,\ldots,k_n).$ We denote
by $h_i\in \mathfrak{H}$ the sum
$$h_i=\sum_{s=1}^n (\Phi_i)_{ss}k_s $$ Then
$H_2^{\infty}(z)=\sum_{i=-\infty,\infty} h_i z^{-i-1}.$ Let us firstly note
that the considered centralizer must commute with $h_1$ and $h_{-1}.$ Let us
consider an infinite series $\sum_{i=-\infty}^{\infty} x_i y_i$ such  that
$x_i\in U(\gg[t^{-1}]),~y_i\in U(t\gg[t])$ which is an element
of the considered completion, i.e. there is a finite number of terms of each
bigrading. The operators $[h_1,*]$ and $[h_{-1},*]$ are homogeneous of
bidegrees $(0,1)$ and $(1,0)$ respectively, hence the question can be
restricted to the original algebra (not completed) and the answer is given
by
\bea
C(h_1)=U(\g[t^{-1}])\otimes \mathfrak{H}_+\qquad
C(h_{-1})=\mathfrak{H}_-\otimes U(t\g^{op}[t])
\nn
\eea
Their intersection in a completed sense is exactly $\mathfrak{H}.$
\\
We have obtained that at the generic point $\hbar$ the commutative
subalgebra $M^\hbar$ belongs to the centralizer of $QI_2$ and in the limit
$\hbar\rightarrow\infty$ generate the centralizer. By the general principle
$M^\hbar$ coincides with the centralizer also at the generic point. The only
thing to do to finish the proof is to recall the Sugawara formula for the quadratic central
elements of $U_{crit}(\widehat{\g})$
$$c_2(z)=:Tr(L^2_{full}(z)):$$
which projects to $QI_2(z)$ up to a central element in $U(\g[t^{-1}])\otimes U(t\g[t])$
$\blacksquare$

\section{Geometric Langlands correspondence \label{Langl-sect}}
Langlands program is one of the most important themes in modern mathematics.
Let us explain how the material of the present paper is related to this program.
The Langlands correspondence has been transferred to  $\CC$-schemes
in \cite{BD}, has
attracted much attention in \cite{L} (see \cite{Frenkel05} for
up-to-date introduction). Recently a connection with the S-duality in gauge theory (\cite{Kap})
 has been discussed \cite{Witten05}
\footnote{A fundamental manuscript deriving the Langlands duality
from the physical arguments appeared recently \cite{WK06}}
(see also \cite{LP}
for the other
relations with physics).
The survey \cite{Frenkel95} is close to the setup of the present
work, it was the source of inspiration for us for years.

\subsection{Local Langlands correspondence over $\CC$ (critical level)}

The aim of the Langlands  correspondence is to relate two type of objects: a representation of
the Galois group of a $1$-dimensional scheme and a representation of the group  $G(\A)$,
 where
$\A$ is the ring of adels and $G$ is a reductive group.
Let us restrict to the case $G=GL_n$.
Consider the local version of the Langlands correspondence over $\CC$ which
means that the scheme is just the  formal punctured
disc $Spec \CC((z)).$
\\
{\bf Local Langlands correspondence} over $\CC$ states the following bijection
\bea
 Reps GL_n(\CC((z))) \rightleftarrows
Morphisms\Bigl(Galois_{Spec(\CC((z)))} \to GL_n(\CC) \Bigr)
\eea
\\
Using the common wisdom one substitutes representations of the
Lie group by representations of the corresponding Lie algebra
and representations of the Galois group by
connections on $Spec \CC((z)).$ One obtains the following {\bf reformulation}
\bea
 Reps U(\widehat{\g}) \rightleftarrows \mbox{Classes of connections on~} Spec \CC((z))
\eea

Naively one expects to parameterize
representations of $U(\widehat{\g})$ by values of central elements
in these representations, it is not true in general but
according to \cite{FeiginFrenkel}  there exists a class
of representations of $U_{crit}(\widehat{\g})$
parameterized by values of Casimirs on them.
This means that for  $\chi: Z(U_{crit}(\widehat{\g}))\to \CC$
one is able to construct $V_{\chi}$ which is a representation
of $U_{crit}(\widehat{\g})$ on which the center acts via this character.
Hence by {\bf restricting to the critical level} and using \cite{FeiginFrenkel}
we have
\bea
\mbox{FF:  } Reps U_{crit}(\widehat{\g}) \rightleftarrows
\Bigl( \chi: Z(U_{crit}(\widehat{\g}))\to \CC \Bigr)
\eea

According to theorem \ref{center} the center of
$U_{crit}(\widehat{\g})$ is isomorphic (as a commutative algebra)
to the commutative subalgebra
$M_{full}\subset U(\g[t^{-1}]\oplus t\g^{op}[t])$ defined with the help of $"det"(L_{full}(z)-\p)$.
This provides the following correspondence
\bea
 \Bigl( \chi: Z(U_{crit}(\widehat{\g}))\to \CC \Bigr)
    \rightleftarrows \Bigl( \chi: M_q^{full}\to \CC \Bigr)
\eea
To obtain a connection on $Spec \CC((z))$ we have to apply the character to
the "quantum characteristic polynomial"
\bea
\Bigl( \chi: M_q\to \CC \Bigr)\nn\\
    \rightleftarrows  \nn\\
~~\chi( "det"(L^{full}(z)-\p))
    =\sum_{k=0}^nC_n^k(-1)^{n-k}\chi[QI_k(z)]\p^{n-k}
\eea

\paragraph{Summary }~\\~
We have the following realization of the correspondence:

\bea
\mbox{ Representation }~~V_{\chi}~~ of~~ U_{crit}(\widehat{\g})
 \nn\\ \rightleftarrows \nn\\
  \mbox{connection on  Spec $\CC((z))$ defined by  the scalar differential operator: }\nn\\
\mbox{}\nn
~~\chi( "det"(L^{full}(z)-\p))=
\sum_{k=0}^nC_n^k(-1)^{n-k}\chi[QI_k(z)]\p^{n-k}
\eea

{\Rem  ~} {\em Agreement with filtration.}
One can see that the construction above respects the natural filtration.
 The algebra $U(\g[t^{-1}]\oplus t\g^{op}[t])$ has the
natural projections to $U(\g[t^{-1}]/t^K \oplus t\g^{op}[t]/t^L).$
Let
$M_{K,L} \subset U(\g[t^{-1}]/t^K \oplus t\g^{op}[t]/t^L)$ be the image of such a projection.
Consider a character $\chi$ of $Z(U_{crit}(\widehat{\g}))\to \CC$
which is the preimage of some character $\chi_{M_{K,L}}$, then
the corresponding $G$-oper can be given by
$\chi_{M_{K,L}}("det"(L_{K,L}(z)-\p))$, where
$L_{K,L}(z)=\sum_{i=-K,...,L} \Phi_{i}/z^{i+1}$.

\paragraph{Global case of the correspondence}~\\~
In the case of the global Langlands correspondence over $\CC$ it was advocated
in \cite{BD}
that representations of the unramified Galois group (i.e. the fundamental group)
are parameterized by Hitchin's D-modules on the moduli stack of vector bundles.
Moreover it was proved that Hitchin's D-modules comes from central elements
$Z(U_{crit}(\widehat{\g}))$.
We will show in the subsequent publications (\cite{CT06-1},\cite{CT06-2})
that one can
apply the construction developed in the present paper
to obtain the global Langlands correspondence for algebraic curves
using the methods developed in \cite{CT03-1},\cite{CT03-2}.

\subsection{Higherdimensional Langlands correspondence (speculation)}
The generalization of  the Langlands correspondence to $d$-dimensional schemes
is extremely complicated question, a little is known about it.
The abelian case - the higher dimensional class field theory
has been developed by \cite{ParshinKato78} (see \cite{HLF00} for survey).
There is only
one paper  known to us
which tries to deal with the nonabelian case: \cite{Kapranov95}.
The paper \cite{GinzburgKapranovVasserot95} deals with the case of surfaces,
but the main conjecture presented therein  is more likely the "deformation" of
$D1$-geometric Langlands
correspondence then  truly $D2$-conjecture, in the same way as the Beauville-Mukai
integrable system is a "deformation" of the Hitchin system.

One can try to formulate the Langlands correspondence  as a correspondence between representations
of $GL(\A)$ where $\A$ is the ring of $d$-dimensional Parshin-Beilinson adels of a
scheme $S$ \cite{adels}
and "$d$-representations" of the Galois group of $S$ (\cite{Kapranov95}).
In the abelian case this formulation is different from Parshin-Kato point of view,
it is dual to it (see last section in \cite{Kapranov95}).
Possibly the most general formulation should state a relation between "$r$-representations"
of $GL(\A)$ and  "$d-r+1$-representations" of the Galois group.
The obvious difficulties with such a formulations are the following:
the notion of a "d-representation" of a group is not well-known at present (at least
to the authors, see however \cite{KapranovNew}).
One can try to substitute the notion of a $d$-representation by the notion
of a flat $d$-connection, which is also not well-known at the moment.

The notion of an abelian $2$-connection is nevertheless
relatively well understood due to the ideas of Brylinsky and Hitchin.
A way to deal with this case
was proposed by P. Severa (see \cite{BC}) who
argued that the Courant algebroid should be treated as
an infinitesimal analogue of a gerbe.
We will try to argue below that this gives useful hints for understanding
the multi-dimensional Langlands correspondence.
Possibly the very complicated
recent theory of \cite{BreenMessing} is relevant to the non-abelian  Langlands correspondence,
another hope is related to the
remarkable constructions developed in \cite{Akhmedov} which possibly will  lead
to the right understanding of the theories of non-abelian $d$-connections and "gauge theories of
nonabelian $d$-forms" \cite{Baez},
 but it is far from being clear at the moment.
Moreover the left-hand-side of the correspondence is also not so clear because
a little is known about the representation theory of groups over $d$-dimensional
local fields (see however \cite{Kaz}).
As one can see in the one-dimensional case it was quite important
to centrally extend the group $GL((z)).$ In $d$-dimensions the natural cocycle
which corresponds to the relevant extension is of degree $d+1$,
so we cannot stay in the realm of groups to make this extension. Possibly
the right thing to do is to make an extension in the category of $A_{\infty}$-algebras,
but after doing this the relevant  notion of a representation is not clear.
(As an alternative way to $A_{\infty}$-algebras one can consider
the language of grouplike monoidal categories which works in the abelian case
\cite{DenisOsip}).
Moreover in the $1$-dimensional geometric Langlands correspondence
the critical level was important, it seems nobody has no idea
what is an analog of this phenomenon in $d$-dimensions.

Possibly some ways round of the difficulties above exist:
by the analogy with the gerbe case one can hope that
to each $d+1$ connection one can attribute a $d$ connection on the loop space,
so iterating this construction one can return back to the case of usual connections
but on some hugely infinite-dimensional space. Instead
of doing a $1$-dimensional extension in the sense of $A_{\infty}$-algebras
one should make an infinite-dimensional central extension, something
like the Mickelson-Faddeev central extension.

Nevertheless we express the dream that an analogue of the Talalaev's formula
$"det"(L(z)-\p)$ also exists in $d$-dimensions and should be
of the following type. The $d$-dim Lax operator is
$$
L_d(z_1,...,z_d)=\sum_{i_1,...,i_d} \Phi_{i_1,...,i_d} z_1^{-(i_1+1)}... z_d^{-(i_d+1)}
$$
By the same arguments as above
we see that $det(L_d(z)-\l)$ defines the center of the Poisson algebra
$S(\g((t_1,..., t_d)))$.
One can hope that there exists a "critical level" such that
the algebra (or $A_\infty$-algebra, or whatever) $U_{crit}(\g((t_1,..., t_d)))$
posses the center of the same size as the center of
$S(\g((t_1,..., t_d)))$,
those characters parameterize not all but  at least some reasonable
class of representations of $U_{crit}(\g((t_1,..., t_d)))$

~\\
{\bf The dream}
One can hope that there exists a formula of the type
\bea & : \widetilde{det} (
L(z) -(\widehat \l)):
=\sum_{i} QI_{i}(z)(\widehat{\l})^i
&\nn
\eea
such that the expressions $QI_{i}(z)$  define central elements in
$Z(U_{crit}(\g((t_1,..., t_d))))$ and each
character of the center   $\chi:Z(U_{crit}(\g((t_1,..., t_d)))) \to \CC$
produces a $d$-G-oper and hence a $d$-connection on $Spec(\CC((z_1,..., z_d)))$
in the following simple way: $\sum_{i} \chi(QI_{i}(z)) (\widehat{\l})^i$
\\
This would be a higher-dimensional analogue of the local Langlands
correspondence.

{\Rem ~} $L(z_1,...,z_d)$ should be thought of as a $d$-form.

{\Rem ~} In the abelian case it seems that the right hand side of the
correspondence is quite clear:
one should consider the expression $\chi(d+L(z))$ as a connection on the Courant algebroid.
The question is what should be on the LHS ? In the $1$-dimensional case
there is a reformulation of the Langlands correspondence: on the LHS
instead of representations of the adelic group one considers sheaves on the Jacobian.
In $2$-dimensions there should be something
analogous and it should be very simple at least for $2$-forms
which have only regular singularities (the case of the tame ramification).

{\Rem ~} From the construction in $d=1$ one can guess that $\widehat \l$ is somehow
analogous to the de-Rham differential $d$. The condition of flatness of the connection
is trivial since any $d$-form on a $d$-scheme is closed.
More precisely one should speak not about $\CC((z_1,..., z_d))$ but
about $d$-dimensional local fields is the sense of Parshin-Beilinson.

\section{Separation of variables}

Let us give some comments why the natural quantization
of the characteristic polynomial $det (L(z) -\l)$ is
$"det" (L(z) -\p)$. The most mysterious part is the appearance
of $\p$ instead of $\l$. In fact this is conceptually based on the idea of
separation variables.

\subsection{Separation of variables for classical systems}

Consider a Poisson algebra $A$ and its maximal Poisson commutative subalgebra $M$
constructed with the help of the Lax operator
with spectral parameter (this means that there exists a $Mat_n\otimes A$-valued function
$L(z)$ such that:
$\forall z,\l$:  $det(L(z)-\l)$  belongs to $M$ and all the elements of $M$
can be expressed as
functions of $det(L(z)-\l) $ for different $z,\l$).

~\\
{\bf Claim (Sklyanin)} { (Classical separation of variables)}
Assume we have a Poisson algebra $A$, its center $Z$, its maximal Poisson commutative
subalgebra $M$
given by the Lax operator $L(z),$ then
there exists such elements $z_i,\l_i$ in $A$ (called separated coordinates) that
\begin{itemize}
\item $\{z_i,\l_j \}=\delta_{ij} f(z_i,\l_j) $ ~~~~~~~ $\{\l_i,\l_j \}=\{z_i,z_j \}=0$
\item $z_i,\l_i$ and $Z$  are algebraically independent and  algebraically generate $A$
\item  $\forall i~$   $det(L(z_i)-\l_i)=0 $.
\end{itemize}
{\Rem~} In fact the variables $z_i,\l_i$ belong not to $A$ but to some algebraic
extension of the field of fractions of $A$,
this means that $z_i,\l_i$ are algebraic functions (not polynomial).

{\Rem~} Sklyanin gave the "magic recipe" to find separated variables:
\\
One should take the {\em properly} normalized Backer-Akhiezer function $\Phi(z,\l)$,
i.e. the solution of $(L(z)-\l)\Phi(z,\l)=0$
and consider the poles $(z_i,\l_i)$ of this function on the corresponding spectral curve.
The obtained set of functions $z_i,\l_i$ is such
that their number is precisely what one needs and the Poisson brackets between them
are separated: $ \{z_i,\l_j \} =\delta_{ij} f(z_i,z_j)$.
The non-algorithmic problem of the method  is to find the proper
normalization of the BA-function but it can be done for a
large class of models (see surveys \cite{Skl92QISM,Skl95SV}).

{\Rem~}This point of view  is usually adopted in modern texts on separation of variables,
it is  related in an obvious way with the classical definition \cite{LLv1}.

{\Rem~} Let us emphasize that the action-angle variables are not the same as separated
ones in general. According to the textbook point of view the
action-angle variables give an example of separated variables,
but not vice versa in general. Modern practice of integrable
systems coins the following crucial difference between these two sets of variables:
we always hope that there are separated variables such that
they can be expressed via algebraic functions of the original variables,
meanwhile in general there is no hope to do this for the action-angle
variables:  even in the simplest example - the harmonic oscillator $H=p^2+q^2$ - the angle
variable $\phi=arctg(p/q)$ is not an algebraic function.

{\Rem~} There exists an alternative (but deeply related) geometric recipe
to find separated coordinates using bihamiltonian structures.
It is being developed by Italian group (F. Magri, G. Falqui, M. Pedroni, et.al.)
\cite{BiHSV}. The recipe characterizes
separated coordinates as Poisson-Nijenhuis coordinates. At the moment the
bihamiltonian method have not been generalized to the quantum level.
It would be a marvelous advance to find a quantum analogue of the Lenard-Magri
chain on the quantum level, this means that starting from the Casimirs on the quantum level,
one should be able to reproduce all the Lenard-Magri chains of the quantum
commuting hamiltonians. One should find a recipe to obtain the formula $"det"(L(z)-\p)$
from the Casimirs defined by the Molev's formula
(\cite{Molev}) $\widetilde{det}(\tilde L(z))$,
such that the classical limit reproduces the classical Lenard-Magri chain
(see \cite{FalquiMusso}).

\subsection{Quantum separation of variables and QCP}

Let us comment on quantum separation of variables and its relation to the
quantum characteristic polynomial along the Sklyanin's ideas.

Let $A,M,Z,L(z)$ be as in the previous subsection.
Denote by $A_q$ the deformation quantization of the algebra $A$, $Z_q$ its center,
$L_q(z)$ - the quantum Lax operator.
{\Conj \label{con-q-SV} (Quantum separation of variables)}
Assume that the bracket in separated variables is  $\{z_i,\l_j\}=\delta_{ij}$.
Then
there exist elements $\hat  z_i, \hat \l_i$ (called separated coordinates)
in $\widehat{A_q}$ -
an algebraic extension of the field of fractions of $A_q$, such that
\begin{itemize}
\item  $"det"(L_q(z)-\p/\hbar)$ generates a commutative subalgebra in $A_q$ which
quantizes the subalgebra $M$
\item $[\hat z_i,\hat \l_j ]=\hbar\delta_{ij}  $
\item $\hat z_i,\hat \l_i$ and $Z_q$  are algebraically independent and
generate $\widehat {A_q}$
\item $\hat z_i,\hat \l_i$ satisfy the quantum characteristic equation
$$\sum_{k=0}^n QI_k(\hat z_i)C_n^k(-1)^{n-k}(\hat \l_i)^{n-k}=0$$
 \end{itemize}
~\\
{\em On definition of $L_q(z)$ from $L(z)$.} In numerous examples the
matrix elements of $L(z)$ are generators of the algebra
$A$ and the algebra $A_q$ can be described as an algebra
with the same set of generators, so typically $L_q(z)=L(z)$.
We usually write $L(z)$ instead of $L_q(z)$.
~\\
{\em Field of fractions.} The construction of the field of fractions
for the noncommutative algebra $A_q$ hopefully can be done in the spirit
of \cite{GK66}. For example in the case of an associative algebra
(not necessarily the universal enveloping algebra)
of less then exponential growth the remark of \cite{KKM}
(section 1.4)  shows that the Ore condition is satisfied.
The way round these problems is to work in the same manner as in
\cite{EnriquezRubtsovSkewFields}.

{\Rem~}
Despite the expected high complexity of the expressions $\hat z_i,\hat
\l_i$ in terms of the original generators in the case of the Gaudin model it is known that
the generators of $M$ can be expressed via $\hat z_i,\hat \l_i$ by relatively simple,
{\em rational} expressions proposed in \cite{Babelon,EnriquezRubtsovSkewFields}.
It is natural to believe that it is a general phenomena.
~\\
{\bf Example} In the context of the present paper one should take
$A=S(\g[t]/t^N)$,\\
 $L(z)=L_q(z)=\sum_i \Phi_i z^{-(i+1)}$,
$A_q=U(\g[t]/t^N),$  $M_q$ is a commutative subalgebra generated by $"det"(L(z)-\p)$.

\subsection{Generalizations of the definition of QCP}
For the case when the Poisson bracket between the separated variables is more
complicated -
$\{z_i,\l_j \}=\delta_{ij} f(z_i,\l_j) ,$ our approach should be modified.
One should quantize this Poisson bracket, define
$\hat z_i,\hat \l_j$ and define an analog of the QCP by:
\bea
"det"(L(\hat z)-\hat \l)=\sum_{k=0}^n QI_k(\hat z)C_n^k(-1)^{n-k}(\hat \l)^{n-k}
\eea
here $\hat z,\hat \l$ are auxiliary variables (not elements of $A_q$)
which satisfy the same commutation relation as quantum separated variables
$\hat z_i,\hat \l_j$.

~\\
Let us describe several examples of the generalized QCP.
In all cases we believe that it will satisfy the expected properties:
\begin{itemize}
\item $"det"(L(\hat z)-\hat \l)$ generates a commutative subalgebra in $A_q$ which
quantizes the subalgebra defined by $det(L( z)- \l)$.
\item Separated variables $\hat z_i,\hat \l_i$ satisfy the quantum characteristic equation
$$\sum_{k=0}^n QI_k(\hat z_i)C_n^k(-1)^{n-k}(\hat \l_i)^{n-k}=0$$
\end{itemize}

\subsubsection{ QCP for the Yangian (XXX-model)}

Consider the case of the Yangian  $Y(\g)$. It is the symmetry algebra of
XXX-spin chain, Toda, DST and Thirring models.
It is well known that the
Poisson bracket between separated variables $z$ and $\lambda$
is given by $\{z,\l\}=\l$.

~\\
The QCP for $Y(\g)$ should be defined by the formula
(see \cite{CT04}: introduction, remark 3):
\bea
QCP^{Yang} = Tr A_n
(e^{\d-\h\pu}T_1(z)-1)(e^{\d-\h\pu}T_2(u)-1)\ldots(e^{\d-\h\pu}T_n(u)-1)
\eea

~\\
It was proved in \cite{Talalaev04}, \cite{CT04} that it produces a commutative subalgebra in Yangian.
It is just the  Bethe subalgebra considered in
\cite{Skl95SV} formula 3.16 and in \cite{NO}.

\subsubsection{ QCP for $U_q(\widehat{\g})$ and elliptic algebras (XXZ,XYZ-models)}

Let us express some {\em preliminary} ideas on the generalization
of the QCP to quantum affine and elliptic algebras.
\\
For the quantum affine algebra $U_q(\widehat{\g})$ the QCP should have the form:
\bea
QCP^{Uq} = Tr A_n^q
(L_1^{+}(\hat z)-\hat \l)(L_2^{+}(z)- \hat \l)\ldots(L_n^{+}(\hat z)-\hat \l)
\eea
where the variables $\hat z, \hat l$ should satisfy a relation
of the type: $\hat z  \hat \l =q \hat \l \hat z$,
$L^{+}(z)$ is the positive current for $U_q(\widehat{\g})$.
 $A_n^q$ should be a version of the $q$-antisymmetrizer
considered in \cite{Arnaudon} Appendix C:
\bea
A_n^q=1/n!\prod_{1\le a < b \le n} R_{ab}^{trig}(q^{a-b})
\eea
This formula generalizes the standard (Cherednik's) formula for the standard
antisymmetrizer in terms of the rational $R$-matrix.
\\
We hope that there is an analogous formula
in the elliptic  case:
\bea
QCP^{ell} = Tr A_n^{ell}
(L_1^{+}(\hat z)-\hat \l)(L_2^{+}(z)- \hat \l)\ldots(L_n^{+}(\hat z)-\hat \l)
\eea
where $A_n^{ell}$ is an appropriate antisymmetrizer,
defined by some analog of the formula above with the elliptic $R$-matrix.
Possibly the facts and methods from \cite{BelJimbo} are relevant to the definition of $A_n^{ell}$.
The  quantum relation
between $\hat z, \hat \l$ can be obtained from the known
Baxter equation.
We hope that with appropriate modifications the QCP
can be defined for different types of elliptic Lax
operators: Sklyanin's like (see e.g. \cite{Zabrodin} formula for $T(\l)$
before formula 5.2), $GL(n)$-analogs of Sklyanin's type Laxes \cite{CZot}
or based on elliptic algebras \cite{EllJimbo}.

\subsubsection{ QCP for the trigonometric and elliptic Gaudin models }

XXZ and XYZ models can be degenerated to the trigonometric
and elliptic Gaudin models.
It was proved in  \cite{SklTakebe} that the Poisson
bracket in terms of the separated variables in these cases is given by $\{ z, \l \}=1$.
So it is natural to guess that the QCP for these
models should be defined in a similar way:

\bea
QCP_{Gaudin} ^{trig, ell} = Tr A_n^{trig, ell}
(L_1^{+}(z)-\p)(L_2^{+}(z)- \p)\ldots(L_n^{+}( z)-\p)
\eea
where $A_n^{trig, ell}$ are the analogs of the antisymmetrizer discussed
in the previous subsection.

\subsection{Universal Baxter equation, $G$-opers and Bethe ansatz}
The Baxter equation (Baxter $T-Q$ relation) and the Baxter $Q$-operator
are considered now as the most powerful  tools to find the spectrum of integrable models.
The equation and the $Q$-operator were introduced in the context of
XYZ model in \cite{Baxter}.
Nowadays studying the Baxter equation and its applications is
an active field of research (see \cite{QOp} and survey \cite{SklyaninQOp}).
However the known constructions are mostly restricted to the cases
of models related to $GL(2),~ GL(3).$
\\
Here we propose a general point of view on the Baxter equation and $Q$-operator;
formulate the conjecture, which relates them to the spectrum of an integrable model;
connect it with the Knizhnik-Zamolodchikov equation and propose a new way
to construct $Q$-operator.
We also explain that G-opers considered in the Langlands correspondence are
closely related to these considerations.
\\
Let $A_q, M_q,  L(z), \hat z,\hat \l$ be as in the previous subsection,
i.e. $A_q$ - an associative algebra (algebra of quantum observables),
$M_q\subset A_q$ - a maximal commutative subalgebra (subalgebra of quantum
commuting hamiltonians) constructed with the help of the QCP $"det"(L(\hat z) -\hat \l)$
for the Lax operator $L(z)$.
Consider $(\pi,V)$ - a representation of the algebra $A_q.$
Let the quantum separated variables $\hat z,\hat \l$  be realized as operators on
the space $\CC((z))$, where $\hat z $ acts as the multiplication by the function $z$,
and $\hat \l$ acts as some differential (difference) operator $\tilde \l$.
\\
{\bf Construction of Baxter equation~}
The differential (difference) equation
$$\pi("det"(L( \tilde z)-\tilde \l)) Q(z)=0$$
for an $End(V)$-valued function $Q(z)$ is called the Baxter equation.
\\
One can
see that for the $SU(2)$-Gaudin, $SU(2)$-XXX, XXZ models this coincides with the degeneration
of the original Baxter equation.
\\
Let us mention that
the basic properties of the $Q$-operator:
\bea
[Q(z),Q(u)]=0, ~~~~~~ [Q(z), m]=0, \forall m\in M_q
\eea
are almost automatic in our approach.
~\\
{\bf Main Conjecture}
Consider  a character $\chi: M_q\to \CC$ of the commutative algebra $M_q$,
consider the scalar differential (difference) equation $\chi ("det"(L(\tilde z)-\tilde \l)) q(z)=0$,
then this equation has trivial monodromy ("$q$-monodromy" i.e. Birkhoff's connection matrix \cite{Birk})
{\bf iff}
there exist a unitary representation $(\pi,V)$ for $A_q$  and a joint
eigenvector $v$ for all commuting hamiltonians $m\in M_q$
such that $\pi(m) v= \chi(m)v$.

~\\
{\em Corrections.} For the models related not to $GL(n)$ but to a
semisimple  group $G$ the  monodromy disappear only after passing to the  Langlands dual group $G^L$.
If irregular singularities appear in $\chi ("det"(L( z)-\tilde \l)) $
one should require the vanishing of the Stokes matrices.

~\\
{\em Simplification for compact groups.} In the case of the $U(n)$-Gaudin and $U(n)$-XXX models
(or in the other words $GL(n)$-models with finite-dimensional representations)
the conjecture can be simplified requesting instead of absence
of monodromy  the rationality of all functions $q(z)$ satisfying the equation above.
In these cases  the conjecture should essentially  follow
(at least in one direction) from the results of
\cite{FMTV} (see also \cite{CT04}).

~\\
The results from \cite{Zabrodin} hint that for the
 $U(n)$-XXZ and  $U(n)$-XYZ there should be some trigonometric
and "elliptic polynomial" solutions of the Baxter equation.
Zabrodin presented a solution of the Baxter equation for the $SL(2)$-XYZ-model
in the form  of elliptic generalization of the hypergeometric series,
the generalizing classical series representation of the Jacobi polynomials.
The general theory of elliptic analogs of hypergeometric functions
developed in \cite{Spiridonov} should be relevant here.

~\\
{\em The case of noncompact groups.}
We are quite sure that the conjecture above is  true (at least for generic $\chi$)
for the case of compact groups (or their q-analogs) but
possibly it should be corrected for the case of noncompact
ones, requiring additional analytical properties for $Q(z)$.
It seems that the situation can be clarified considering
as an example the closed Toda chain, which is connected with the
representation of the noncompact form of the Yangian for $sl(2).$ It is
interesting to
compare our approach with the classical results due to
Gutzwiller, Sklyanin, Pasquier, Gaudin (see \cite{Kharhev}).

~\\
{\em Relation to G-opers.}
For the case $A_q=U(\g[t])$ the differential operator\\ $\chi ("det"(L( z)-\tilde \l))$
is precisely the $G$-oper which was considered in the Langlands correspondence.

~\\
{\em Relation to Bethe equations.}
It is known that the condition of absence of monodromy can be
efficiently written down as the Bethe equations in the case of the Gaudin and XXX models.
In general an analogous procedure is not known.
However for example in the case of regular connections
the necessary condition for absence of monodromy
is the very trivial condition for the  residues  of
the connection to have integer spectrum.
Hopefully the Bethe equations for the spectrum is
the consequence of the similar local condition and
by some unknown reasons the absence of local
monodromies guarantees the  absence of the global one.

~\\
{\em "Bethe roots" are zeros of the function $q(z)$.}
Let us give an example for the Gaudin model related to $SU(2)$,
taken from \cite{Frenkel95} section 5.4.
Consider the Hilbert space\\ $V_1\otimes ... \otimes V_N$, where
$V_i$ are finite-dimensional irreps of $sl(2)$ of weights $\l_i$.
Considering a character $\chi$ we see that the equation
$\chi( "det"(L(z)-\p))$ takes the form:
\bea
\chi( "det"(L(z)-\p))=\p^2-\sum_i \chi(C_i)/(z-z_i)^2-\chi(H_i)/(z-z_i)=\nn\\
=\p^2-\sum_i (\l_i(\l_i+2)) /4(z-z_i)^2-(\mu_i)/(z-z_i)
\eea
where $C_i$ - are Casimirs of the $i$-th copy of $sl(2)$ and $H_i$ are Gaudin hamiltonians.
\\
One can see that the Bethe equations:
\bea
\sum_i \l_i/(w_j-z_i)- \sum_s 2/(w_j-w_s)=0
\eea
are equivalent to the condition that the function
$q(z)= \prod_i^N (z-z_i)^{-\l_i/2}  \prod_j (z-w_j)$
satisfies the equation:
\bea
(\p^2-\sum_i (\l_i(\l_i+2)) /4(z-z_i)^2-(\mu_i)/(z-z_i))q(z)=0
\eea
\\
One sees that the Bethe roots $w_i$ arise as zeros of the function $q(z)$
which seems to be a general phenomenon.

~\\
{\em Explanation by separation of variables.}
The best way to explain the origin of this conjecture
is from the point of view of the separation of variables.
Consider the quantum separated variables $\hat z_i, \hat \l_i$, then representations of $A_q$
can be realized in the space of functions of $z_i$ where $\hat \l_i, \hat z_i$
act as operators $\tilde z_i, \tilde \l_i$ and Casimir elements act by some constants.
From the basic identity  $"det"(L(\hat z_i)-\hat \l_i)=0 $ we see that
joint eigen-vectors for $m\in M_q$ can be found in a factorized form:
$$\Psi(z_1,\dots,z_n)=\Psi^{sv}(z_1)\dots\Psi^{sv}(z_n)$$
where $\Psi^{sv}(z)$ satisfies the
equation $\chi( "det"(L(\tilde z)-\tilde \l)) \Psi(z)=0$.
By the standard quantum-mechanical intuition \cite{LLv3}
we see that this solution belongs to the physical Hilbert space iff
it is single valued, i.e. have trivial monodromy.

~\\
Let us formulate the important problem which should be of  interest not only for experts
in mathematical physics but also  in pure representation theory.
~\\
{\bf Problem~} To construct a universal Baxter $Q$-operator directly
in the algebra $A_q$, i.e. to construct an $A_q$-valued
function $Q(z)$ which satisfies the "universal Baxter equation":
$$("det"(L( \tilde z)-\tilde \l)) Q(z)=0$$
\\
{\bf Approach~ }
The most important profit of our approach 
shown in  \cite{CT04} (see also sections below) is the following:
$Q(z)$ can be constructed
 from $\Psi(z)$ - the fundamental solution
of the equation $$ (\tilde \l-L_q( \tilde z))\Psi(z)=0$$ which is
the  universal analogue of the KZ-equation.
In our paper the case of the Gaudin and XXX models was considered but
we believe that it is a general phenomenon.


{\Rem ~ \label{DoreyTateo98Rem} } In \cite{DoreyTateo98}
amusing observations were made:
in the simplest case they relate the regularized characteristic
polynomial of Schrodinger operator $H=(\partial_{x}^2+x^4)$
(called spectral determinant and defined as $D(E)=C_0\prod_{i} (1+E/E_i)$,
where $C_0$ is some normalizing constant,  $E_i$ - eigenvalues of $H$)
to the Baxter $T-Q$ relation for some quantum integrable model.
They observe the coincidence with the relation from \cite{Voros}:\\
$$D(E \epsilon^{-1}) D(E) D(\epsilon E) = D(E\epsilon^{-1}) +  D(E) +D(E\epsilon) +2$$
where $\epsilon=exp(2\pi ~i/3)$, with the relation for some quantity
related to the ground state energy in the theory of perturbed parafermions.
Moreover they found similar relations between spectral determinants
of $H=(\partial_{x}^2+x^\alpha)$ and $Q$-operator's vacuum eigen-values
in the conformal field theories.
It is tempting to ask  a question is there any analogous
relations for the Gaudin model ?
This is possibly  related to the
bispectrality \cite{DualIS}
which often occurs in integrable systems,
this means that a joint eigen-function of the commuting
hamiltonians $H_x \Psi(x,\l)=h(\l) \Psi(x,\l)$
often satisfy some dual set of equations
$ H^D_{\l} \Psi(x,\l)=h^D(x) \Psi(x,\l).$

\section{ KZ equation }
\subsection{KZ equation via the Lax operator and rational solutions}

Let us emphasize the connection of the material above with the KZ-equation
which is of somewhat different  (but related) nature then the connection of KZ-equation
and Gaudin model  known before (see \cite{KZ-Gaudin}).
\\
Denote by $e_{kl}$ the standard basis in $\g$.
Recall that the Lax operator for the Gaudin model (see formula \ref{main-part-Lax-Gaud})
for the case $K=0$ is given by:
\bea
L^G(z)=\sum_{i=1...N}
\frac{\Phi^G_{i}}{z-z_j}=\sum_{i=1...N}  \sum_{kl}
\frac{E_{kl}\otimes e_{kl}^{(i)}}{z-z_i}
\eea
\\
Recall the definition of the KZ equation \cite{KZ} for $\g.$
Let  $z_i\in\CC,~$ $(\pi_i,V_i)$ be arbitrary representations of $\g$.
The KZ-equations is a system of ordinary differential equations for
a $V_0\otimes ... \otimes V_N$-valued
function $\Psi(z_0,...,z_N)$ given by:
\bea
\label{KZ}
(\partial_{z_j} +\sum_{i\ne j} \sum_{kl} \frac{\pi_j(e_{kl})\otimes
\pi_i(e_{lk}) }{z_j-z_i}) \Psi(z_0,...,z_N)=0
\eea
~\\
Let $V_0=\CC^n$ be the antifundamental
representation i.e. $\pi_0(e_{kl})=-E_{lk}\in Mat_n$.
\\
{\bf Observation} The expression
$$(\pi_1\otimes ... \otimes \pi_N) (\p -L^G(z))$$
coincides precisely with one of the KZ-operators
$$\partial_{z_0} +\sum_{i\ne j} \sum_{kl} \frac{\pi_0(e_{kl})\otimes \pi_i(e_{lk})
}{z_0-z_i}$$
under the agreement $z=z_0.$
\\
Let us introduce a degenerated version of the KZ-equation which is natural from the point
of view of the present paper.
Consider $L(z)=\sum_{i=0,...,N} \Phi_i/z^{i+1}$ the Lax operator for $U(\g[t]/t^N)$ and
an arbitrary finite-dimensional representation $\pi$ of $U(\g[t]/t^N)$
in a vector space $V.$

{\Def ~ } Let us call the degenerated KZ-equation for a $V$-valued function $\Psi(z)$
the following one:
\bea
\label{KZ-degen}
(\p -\pi L(z))\Psi(z)=0
\eea
\\
In view of the connection with Bethe ansatz theory and the Baxter equation  (see next subsection)
the following conjecture is important:
{\Conj \label{KZ-conj}
For the case of finite dimensional representations
the KZ-equation \ref{KZ} and its degeneration \ref{KZ-degen} has only rational solutions.
}

{\Rem ~}
We have tested
the conjecture above in some simple
examples with the help of "Mathematica" and obtained the rational solutions.

{\Rem ~} We can conjecture even that
all the solutions of the  equations
$$(\p -k\pi L(z))\Psi(z)=0~~\mbox{ and~~} (\p -k\pi L^G(z))\Psi(z)=0$$
are rational functions for $k$ integer.

{\Rem ~} Our conjecture can be considered as a special limit of
the Kohno-Drinfeld theorem, which says that the monodromy representation
of the KZ-equation is equivalent to the representation of the braiding group
defined
by the  $R$-matrix of the corresponding quantum group. This quantum $R$-matrix becomes
an identity matrix for our case.
The problem is that the proof of this theorem works only for generic
values of $k$, the integer values cannot be treated in such a simple manner.

{\Rem ~} Another support for the conjecture comes
from the Matsuo-Cherednik correspondence \cite{Cherednink-Matsuo},
which relates solutions of the modified KZ-equation to solutions of the
Calogero model, by another hand it is well-known that solutions of the Calogero model
for integer values of $k$ are polynomials. The obstacle to apply directly this result to our
problem is that
when the relevant parameter $\l$ tends to zero the modified KZ equations
degenerate not to the usual KZ-equations (as one can naively expect),
but to slightly different equations
(\cite{Cheredniklectures},\cite{FelderVeselov}).
We are indebted to A. Veselov for discussion on this point.

{\Rem ~} Also we were informed by A.Veselov about his common unpublished result
with G.Felder which gives explicit formulas for the
rational KZ solutions for the $GL_n$ case if one
considers representations in $\CC^n$.

{\Rem ~} For the case of the modified KZ-equation of the form:\\
 $$ (\p -k\pi L^G_K(z))\Psi(z)=0 $$
where $L^G_K(z)$ is the Gaudin Lax operator with a constant
term $K$ one should expect the appearance of non-polynomial solutions which
nevertheless have no monodromy.

{\Rem ~} For the KZ-equation related to
semisimple groups one should expect mild monodromy which
disappears after passing to the Langlands dual group.

{\Rem ~} Interesting polynomial solutions of the KZ and q-KZ equations were found in
\cite{Pasq}, \cite{ZinJust} but their situation is different from ours - the value $1/k$ is integer,
there is no rational fundamental solution. Their solution
is somehow related to the
conjecture \cite{RazStrog} on the ground-state of the $O(1)$ loop  model
 and to the quantum Hall effect. We are indebted to P. Zinn-Justin for
discussions on this point.

\subsection{Generalized KZ-equations}%
The ideas presented above force us to introduce
generalized versions of the KZ-equation, related to models other
then the Gaudin model.

Let us take $A_q$, $L(z)$, $\hat \l, \hat z$
 as in previous subsections, i.e.
$A_q$ - an associative algebra (algebra of quantum observables),
$\hat \l, \hat z$ - auxiliary variables with the same
commutation relation as for separated variables, $\hat z_i, \hat \l_i$ -
quantum separated variables.
Let the operators $\tilde z, \tilde \l$ provide a representation of
 $\hat \l, \hat z$ in the space $\CC((z))$.

As an example, alternative to the Gaudin model,
one can consider $A_q$ - the Yangian $Y(\g)$,
then $\tilde z$ is an operator of multiplication by $z$,
$\tilde \l = exp ( \p)$.
Another example is the quantum affine algebra $A_q=U_q(\g)$,
here $\tilde z$ is an operator of multiplication by $exp(z)$,
$\tilde \l = exp ( ln(q) \p)$.
Other examples can be defined from the elliptic algebra,
elliptic Gaudin model, the versions of the models above related to semisimple groups.

~\\
{\bf Notations}
Let $(\pi,V)$ be  a representation of $A_q$,
let us call the equation on an $A_q\otimes \CC^n$-valued function $\Psi(z)$ of the type
$
( \tilde \l -L(\tilde z)) \Psi(z)=0
$
the {\em universal generalized KZ-equation}.
~\\
Let us call the equation on a $V\otimes \CC^n$-valued function $\Psi_\pi(z)$ of the type:\\
$
(\tilde \l -\pi(L(\tilde z))) \Psi_\pi(z)=0
$
the {\em generalized KZ-equation}.

{\Conj (Generalized KZ rationality conjecture)}
Let $(\pi,V)$ be a unitary representation of $A_q$,
consider the differential (difference) equation $(\tilde \l - \pi (L(\tilde z))) \Psi_\pi(z)=0$
for a $V\otimes \CC^n$-valued function $\Psi_\pi(z)$.
Then this equation has trivial monodromy 
("$q$-monodromy" i.e. Birkhoff's connection matrix \cite{Birk}).

~\\
{\em Corrections.} For the models related not to $GL(n)$ but to
a semisimple  group $G$ the monodromy disappears only after passing to the Langlands dual group $G^L$.
If irregular singularities appear in $\chi ("det"(L( z)-\tilde \l)) $
one should require the vanishing of the Stokes matrices.
~\\
{\em Relations to q-KZs.} The $q$-analogs of the KZ equations were defined in
\cite{FrRes}, \cite{Smir}, it would be interesting
to find relations.

\subsection{Baxter equation and Q-operator from KZ equation}

Let us preserve the notations of the previous sections.
As it was argued above the equation $\pi("det"(L( z)-\tilde \l)) Q(z)=0$
should be considered as the generalized Baxter equation,
on the other hand the equation $((L( \tilde z)-\tilde \l)) \Psi(z)=0$
can be considered as the generalized KZ-equation.

The main message of \cite{CT04} was the following:  one is able to construct
solutions of the Baxter equation from solutions of the KZ-equation.
This goes as follows:
let $\Psi(z)$ be a solution of the equation
$
(\tilde \l -L_q(z)) \Psi(z)=0
$.
Then any  component $\Psi_i(z)$ of the vector $\Psi(z)$
provides a solution of the universal Baxter equation:
$"det"(L( z)-\tilde \l)) \Psi_i(z)=0$.
\footnote{Moreover, in \cite{CT06-03} we show that  there is a gauge
transformation of the KZ-type connection $(\tilde \l -L_q(z))$
to the Drinfeld-Sokolov form corresponding to the differential (difference)
operator $"det"(L( z)-\tilde \l))$.}

The construction above allows to deduce  (at least in one direction)
the main conjecture 
on the Baxter equation from the KZ
rationality conjecture \cite{CT04}.
Also it gives a hope to apply profound theory
of the KZ equation and Cherednik algebras
to solve the Baxter equation.

\section{Other questions \label{Disc-sect}}

\subsection{Generalization to semisimple Lie algebras}

The construction of a commutative family for a semisimple Lie algebra $\gg$
presumably can be obtained precisely by the same formula:
\bea
"det"(L^{\gg}(z)-\p)
\eea
where $L^{\gg}(z)$ is defined as follows: consider the quadratic Casimir element
$C\in U(\gg)$, define
\bea \Phi^{\gg}=1/2\pi_1(\triangle(C)-C\otimes 1-1 \otimes C)\in Mat_n\otimes U(\gg)
\eea
where $\triangle(C)$ is the standard coproduct for $U(\gg)$, $\pi_1$ is an arbitrary
$n$-dimensional
representation for $\gg$. Define $L^{\gg}(z)$ for the algebra $U(\gg[t]/t^N)$ by the
formulas $L^{\gg}(z)=\sum_i \Phi t^i z^{-(i+1)}$.

It is likely that this construction works only for ADE Lie algebras, for the
others
one should possibly introduce $\Phi^{\gg}$ which incorporates
$\gg$ and the Langlands dual $\gg^L$.
This construction possibly gives not always maximal subalgebras but extendable to maximal
by some expression like $\mbox{\bf Pfaffian}(L^g(z))$.

All the conjectures presented above should also be true in this more general situation,
with the only difference that speaking about monodromy of the KZ-equation and
$G$-opers
one expects that it became trivial only after passing to the Langlands dual group.

\subsection{Generalization to quantum groups}

As we already discussed it is natural to believe that
the construction of the QCP can be generalized to the case of the
Yangians, quantum affine algebras and elliptic algebras.
As we already stated  we also hope the other results presented here
(the separation of variables, the construction
of the Baxter equation and its relation to the spectrum, relation between
KZ and Baxter equations)
should be true in this more general setting.

Let us also remark that hopefully our approach
produces not only the center at the critical level but also the deformed $W$-algebras
out of the critical level.
Following \cite{SemR},\cite{Semenov97} one should introduce the properly
defined full current $L_{full}=L_+(z)L_-^{-1}(z)$, then we hope that
\bea
:QCP(L(z)):=:"det"(L_{full}(\hat z) -\hat \l):
\eea
defines the center at the critical level and the deformed $W$-algebras out of the critical level
for the centrally extended Yangian double \cite{KT},\cite{K}, \cite{Iohara},
quantum affine algebras $U_q(\hat g)$ and
possibly for elliptic algebras also. In the case of quantum affine algebras $U_q(\hat g)$
this deformed $W$-algebras should coincide with the ones introduced in \cite{W}.
The center of quantum affine algebras was described in completely different way in
\cite{SemR},\cite{DE} it would be interesting to clarify the connection.

~\\
The $:...:$ should be understood in the sense of the AKS map $I:A_q^+\otimes A_q^-\to A_q$
given by $I(a\otimes b)=as(b)$, see \cite{Semenov97}.

It would be also interesting to explore an analogue of the approach above
in the case of Lie superalgebras and their $q$-deformations.

\subsection{Application to other integrable systems}

There are several systems  related to the Gaudin model:
the Calogero-Moser system,  tops, the Neumann model,
generalized bending flows. Our approach should be  applicable for them also.

\paragraph{Calogero-Moser system}
Rational and trigonometric Calogero-Moser system with or without spin
can be obtained by  reduction from the system whose Lax operator
satisfies precisely the commutation relation of the Gaudin model:
\bea
[\one L(z) , \two L(u) ] = [ \frac{P}{z-u}, \one L(z)  + \two L(u)  ]
\eea
In the case of the trigonometric model the Lax operator is given
(see \cite{CT03-1}, \cite{CT06-1} for details)
by:
\bea
L(z)=\Phi^L/(z-a) + \Phi^R/(z-b) +\Phi^{orbit}/(z-c)
\eea
where the {\em non}-reduced phase space is $T^*GL(n)\times Orbit$,
$\Phi^L,\Phi^R$ are the elements  related to the left and right
invariant vector fields on $GL(n)$, $\Phi^{orbit}$ is the element of the coadjoint orbit.
The hamiltonian reduction is proceeded with respect to the action of $GL(n)$
by conjugation on $T^*GL(n)$ and the natural action on the coadjoint orbit.
The system obtained by the classical reduction is the trigonometric Calogero-Moser
system.

Since it is generally believed that the reduction and the quantization
commute it is expected that by elaborating the reduction of the Gaudin-type model
and using our results related to it one can obtain:
\begin{itemize}
\item construction of quantum commuting hamiltonians
\item construction of the Baxter equation and Q-operator
\end{itemize}

\paragraph{Classical and Quantum tops}

In \cite{ReTop} one can find the description of multidimensional
tops and the discussion of their quantization with the help
of the center of the affine algebra  at the critical level.
Our results give a direct way to quantize the Lax operators described there
and approach the problem of construction of the $Q$-operator and Baxter equation.

\paragraph{Neumann system.}
Neumann system is a special case of the $SL(2,\RR)$-Gaudin model
related to special unitary representations of $SL(2,\RR)$.
In beautiful papers \cite{TalonNeum},\cite{BT2} the separation of variables
and construction of the Baxter equation were found for this model.

Let us advertise that our approach based on the relation between
the KZ-equation and the Baxter equation can allow one to find
explicit solutions for the Baxter equation using the solutions
of the KZ-equation for $SL(2,\RR)$.

\paragraph{Bending flows}
Bending flows is an amusingly simple and nice integrable system
discovered in \cite{Milson}
(see also \cite{Rag}, \cite{FM01}).
The generalization to the case of $SL(n)$
was proposed in \cite{FalquiMusso2}.
The peculiarity of the proposal is to introduce
not only one Lax operator but the set of Laxes:
$L_a(z)=\sum_{i=1,...,a-1} \Phi_i +z \Phi_a$.
Classical commuting hamiltonians were defined as $det(L_a(z)-\l)$.

We propose that {\em quantum commuting bending flows} can be defined
in the same way as above: $"det"(L_a(z)/(z(z-1))-\p)$.
It follows from the theorems above that for each $a=1,...,n$ the
quantum hamiltonians defined in this way commute,
but one needs to prove the commutativity for
different $a$, we hope to elaborate this in future.

\subsection{Hamilton-Cayley identity
and "Quantum eigenvalues"}

In the subsequent publication \cite{CT06-03} we will describe
the generalization of Hamilton-Cayley identity
for the quantum Lax operator with spectral parameter of the Gaudin
and related models (i.e. for $\g[t]$ Lie algebra).
In this generalization
the quantum characteristic polynomial plays the role
of the standard characteristic polynomial in the
classical Hamilton-Caley identity.
For the Lax operators without spectral parameters
the question has been deeply explored in \cite{HC}.
Some physical applications of classical Hamilton-Caley identity
can be found in \cite{BerensteinUrrutia93}.
Let us mention that  in \cite{GelfandBig} section 8.6 page 96
the generalization of the Hamilton-Cayley identity
was found for the matrix with the coefficients in arbitrary
noncommutative algebra, but this generalization
is of somewhat different nature - the coefficients
of their identity are given not by scalars,
but by the diagonal matrices with (generically) pairwise distinct entries
(see for example formula 166). Nevertheless it is
quite possible that there exists a connection
between these constructions.

~\\
Another related question is to find a factorization of the quantum
characteristic polynomial i.e. to define
"quantum eigenvalues". This question seems to be deeply related to quantum separated
variables.
In \cite{GS}
another version of the characteristic polynomial for the
$GL(2)$ Lax operator was factorized explicitly,
hopefully similar ideas related to the notion of quantum bundles should work in our case.

~\\
Also the paper \cite{MolevRetakh} hints that there is possibly non-trivial
intersection between our approach and the theory created in \cite{Gelfand},
so hopefully their general theory can be applied to our particular cases.
\\
Let us also mention that the questions dealt  in \cite{Kir},
\cite{Rozhkovskaya} seems to be very close to the ones here.

\subsection{Matrix models}
It was pointed out to us by A. Alexandrov that in \cite{AMM} it
appeared the "quasi-classical" curve of the type $y^2=\sum_i H_ix^i$,
where $H_i$ are operators. It is related to a completely different field
  - the genus zero part of the partition function
of matrix models was expressed with its help. So it is natural to
speculate that all genus partition function of the model is related
to the solution of the equation given by the "quantum spectral curve":
$(\partial_x^2-\sum_i H_ix^i)\psi(x)=0$.
It always happens in matrix models and Gromov-Witten theory
that all genus partition function is in a sense a quantization of the genus zero part.

\subsection{Relation with the Drinfeld-Sokolov reduction}

The result of Drinfeld-Sokolov
is that the hamiltonian reduction $\hat{ \g}^* //n_+$ produces naturally
the space of differential operators.
This plays a crucial role in the construction of the
Langlands correspondence. The quantum characteristic polynomial
$"det"(L(z)-\p)$ gives a differential operator from more
or less the same data and plays a similar role in the Langlands
correspondence, so we have no doubts that these two constructions
are deeply connected. It would be very interesting to clarify the relation.

\appendix
\section{Gaudin model with constant term}
\label{LK}
The crucial point in the quantization for the standard Gaudin model (see
\cite{Talalaev04}) is the
construction of the commutative family in $Y(\g)$ due to \cite{Molev}.
Actually the generators
$$
\tau_k(u,\h)=Tr A_n T_1(u,\h)T_2(u-\h,\h)\dots
T_k(u-\h(k-1),\h)C_{k+1}\dots C_n \quad k=1,\ldots n
$$
for a constant
matrix $C$ commute (here the matrix $C$ does not depend on the
spectral parameter $u$).
The Gaudin
model without constant term was quantized with the choice $C=1.$ To add a
constant term one has to consider the matrix $C$ of the form
$$C=1-\h K$$
Composing now the Yangian-type universal $G$-oper of the form
\bea
Qchar(u,\h)&=& Tr A_n
(e^{\d-\h\p}T_1(u,\h)-(1-\h K))\ldots
(e^{\d-\h\p}T_n(u,\h)-(1-\h K))\nn\\
&=&\sum_{j=0}^n\tau_j(u-\h,\h)(-1)^{n-j}C_n^j e^{\d-j\h\p} \label{cr}
\eea
where the generators $\tau_j$ correspond to the choice $C=1-\h K$
one obtains
\bea
Qchar(u,\h)=\h^n qchar(L^K(u)-\p) + O(\h^{n+1})\nn
\eea
where
\bea
L^K(u)=K+\sum_i\frac {\Phi_i}{u-z_i}
\eea

\section{Lax operator with trivial residue}
\label{-2}
One of the natural geometrical
conditions for the Lax operator is the trivial residue at a pole which correspond to the
holomorphity condition for the Lax operator on a singular curve or to the
moment map condition subject to the hamiltonian reduction. Despite the Lax operator
of the form
\bea
\tilde{L}(z)=\sum_{i=1}^N\frac {\Phi_i}{z^{i+1}}
\eea
does not satisfy the $r$-matrix linear commutation relation with
$r(z)=P z^{-1}$ one can quantize the corresponding system by the same
formula as for the complete Lax operator $L(z)$ given by the formula
${L}(z)=\sum_{i=0}^N{\Phi_i}/{z^{i+1}}$.
\\
Let us introduce the positive Lax operator
\bea
\label{L+}
L^+(z)=\sum_{i=1}^N {\Phi_i}{z^{i-1}}
\eea
$L^+(z)$ satisfy the following relation
\bea
[\one L^+(z),\two L^+(u)]=-[\frac P {z-u},\one L^+(z)+\two L^+(u)]
\eea
\\
By theorem \ref{most-general-formulation} the quantum characteristic polynomial
constructed with the help of $L^+(z)$
\bea
qchar(-L^+(z)-\p)=Tr A_n(-L^+_1(z)-\p)\dots(-L^+_n(z)-\p)
\eea
has
commutative coefficients.
\\
On the other hand the transformation
$$L^+(z)\mapsto -\frac 1 {z^2} L^+(\frac 1 z)=\tilde{L}(z)$$
maps the space of positive
Lax operators to the space of operators of the form $\tilde{L}(z).$
\\
Introducing the variable $u=z^{-1}$ we obtain
\bea
\label{ex}
qchar(-L^+(z)-\p)=(-1)^n Tr A_n u^2(\tilde{L}_1(u)-\partial_u)\dots
u^2(\tilde{L}_n(u)-\partial_u)
\eea
Let us introduce the partial characteristic polynomials
\bea
qchar_k(\tilde{L}(u)-\partial_u)=Tr A_n(\tilde{L}_1(u)-\partial_u)\dots(\tilde{L}_k(u)-
\partial_u)
\label{part-qchar}
\eea
and recall the combinatorial result from \cite{CT04} (Theorem 1) that the coefficients
of these expressions coincide with the coefficients of the complete
characteristic polynomial up to constant scalars:
\\
if one introduces $QI_i$ as coefficients of
\bea
qchar(\tilde{L}(u)-\partial_u)=\sum_{i=0}^n
C_n^i(-1)^{n-i}QI_i(u)\partial_u^{n-i}\nn
\eea
then
\bea
qchar_k(\tilde{L}(u)-\partial_u)=\sum_{i=0}^k
C_k^i(-1)^{k-i}QI_i(u)\partial_u^{k-i}\nn
\eea
Now  putting all expressions $u^2$ into the formula \ref{ex} to the left
one obtains
\bea
qchar(-L^+(z)-\p)&=&\sum_{k=0}^n f_k(u)
qchar_k(\tilde{L}(u)-\partial_u)\nn\\
&=&
\sum_{j=0}^n\left(\sum_{i=j}^n QI_{i-j}(u)C_i^j(-1)^j
f_i(u)\right)\partial_u^j
\eea
for some polynomials $f_k(u).$
By construction this is  a differential operator with commuting coefficients.
To finish the proof one needs to verify that the linear transformation from this
coefficients to $QI_i(u)$ is not degenerate over the field of meromorphic
functions. It is true  because the matrix is triangular with $f_n(u)$ on
the diagonal. This function is equal $(-1)^n u^{2n}$ due to the higher order
term count
and hence is not identically zero.

\section{AKS lemmas}
Let $\gg=\gg_+\oplus \gg_-$ be a finite dimensional Lie algebra which
is a direct sum of two its Lie subalgebras. One has an isomorphism
of linear spaces related to some normal ordering
$$\phi:U(\gg) \rightarrow U(\gg_+)\otimes U(\gg_-)$$
{\Lem The center $Z\subset U(\gg)$ maps to a commutative subalgebra
in $U(\gg_+)\otimes U(\gg_-^{op})$ via $\phi$ where
$\gg_-^{op}$ is $\gg_-$ with inverted bracket.
}
\\
{\bf Proof~} Let us denote the commutator in $U(\gg_+)\otimes U(\gg_-^{op})$
by $[*,*]_R.$
Let $c_1,c_2$ be two central elements in $U(\gg)$ taken in the
form
\bea
c_i=\sum_j x_j^{(i)}y_j^{(i)}\qquad x_j^{(i)}\in U(\gg_+),~y_j^{(i)}\in U(\gg_-)\nn
\eea
\bea
[\phi(c_1),\phi(c_2)]_R=[\sum_j x_j^{(1)}y_j^{(1)},\sum_k x_k^{(2)}y_k^{(2)}]_R=
\sum_{j,k}[x_j^{(1)},x_k^{(2)}]_R
y_j^{(1)}y_k^{(2)}+x_j^{(1)}x_k^{(2)}[y_j^{(1)},y_k^{(2)}]_R\nn
\eea
Due to the definition of the algebraic structure
\bea
[x_j^{(1)},x_k^{(2)}]_R=[x_j^{(1)},x_k^{(2)}]\qquad
[y_j^{(1)},y_k^{(2)}]_R=-[y_j^{(1)},y_k^{(2)}]\nn
\eea
\bea
[\phi(c_1),\phi(c_2)]_R=\sum_k [c_1,x_k^{(2)}]y_k^{(2)}-\sum_j
x_j^{(1)}[y_j^{(1)},c_2]\nn
\eea
which is zero due to the centrality of $c_1,c_2~\square$

{\Rem~} The result is valid in the classical case if one consider the Poisson algebra
$S(\gg)$ and the following isomorphisms of linear spaces
\bea
\phi_{cl}&:&S(\gg)\rightarrow S(\gg_+)\otimes S(\gg_-^{op})\nn
\eea

\end{document}